\documentclass[a4paper,fleqn,usenatbib]{mnras}

\usepackage{newtxtext,newtxmath}
\usepackage{mathtools}
\usepackage{cuted}
\usepackage{widetext}
\usepackage[T1]{fontenc}
\usepackage{ae,aecompl}

\usepackage{graphicx}	% Including figure files
\usepackage{amsmath}	% Advanced maths commands
\usepackage[]{algorithm2e}
\usepackage[caption=false]{subfig}
\graphicspath{{figures/}}

\DeclareMathOperator{\atantwo}{atan2}

\title[Quadrupole spin dynamics]{Orbital spin dynamics of a millisecond pulsar around a massive black hole with an general mass quadrupole}

\author[T. Kimpson et al.]{
Tom Kimpson,$^{1}$\thanks{E-mail: tom.kimpson.16@ucl.ac.uk}
Kinwah Wu ,$^{1}$
and Silvia Zane $^{1}$
\\
$^{1}$ Mullard Space Science Laboratory, University College London. Holmbury St. Mary, Dorking, Surrey, RH5 6NT, UK
}

\date{Accepted XXX. Received YYY; in original form ZZZ}

\pubyear{2020}

\begin{document}
\label{firstpage}
\pagerange{\pageref{firstpage}--\pageref{lastpage}}
\maketitle

\begin{abstract}
We investigate the spin dynamics of a millisecond pulsar (MSP) in compact orbit around a Kerr-like massive black hole with an general mass quadrupole. We use the Mathisson-Papetrou-Dixon formulation to compute the orbital and spin evolution of the MSP, accounting for the non-linear interaction of the pulsar's energy-momentum tensor on the background spacetime metric. We investigate how the MSP spin and BH quadrupole moment manifest in the pulsar spin-orbital dynamics. We discuss the astrophysical observational implications of these spin and orbital dynamics on the timing of a radio pulsar in an Extreme Mass Ratio Binary, e.g. a Galactic Centre pulsar. In particular, notable timing variations in the Einstein delay and Roemer delay are observed, along with modifications to the pulsar pulse profile. 
\end{abstract}


%%%%%%%%%%%%%%%%%%%%%%%%%%%%%%%%%%%%%%%%%%%%%%%%%%

%%%%%%%%%%%%%%%%% BODY OF PAPER %%%%%%%%%%%%%%%%%%

\section{Introduction}
The existence of astrophysical black holes (BHs) is well evidenced by gravitational wave astronomy \citep[e.g. LIGO/Virgo observations of binary BH mergers,][]{AbbotBHb} and observations of the BH `shadow' via Very Long Baseline Interferometry \citep[e.g. Event Horizon Telescope observations of the centre of M87, ][]{EHT}. These observations also provide the first tests of GR in the gravitational strong-field, complementing previous successful tests in the solar system \citep{Will2014} and in binary pulsar systems \citep{Lorimer2008}. In spite of this success, there remain open questions on the nature of astrophysical black holes and the validity of Einsteinian GR. These include issues related to the interaction of spin in a curved spacetime \citep{Iorio2012,Plyatsko2016}, the non-uniqueness of the Einstein field equations \citep{Psaltis2008}, the presence of singularities and whether astrophysical BHs are indeed described by the GR solution. \newline

\noindent Millisecond pulsars in orbit around a massive ($10^3 - 10^6 M_{\odot}$) BH are particularly useful systems for the study of strong-field spin dynamics. The remarkable gyroscopic stability of MSPs \citep{Verbiest2009} allow high precision radio timing observations to explore their spin-orbital dynamics, whilst the mass ratio enables a perturbative, mathematically tractable treatment in which the spin dynamically interacts with the background, curved, stationary spacetime (spin-curvature coupling). Such systems in which a MSP orbits a much more massive companion are known as Extreme/Intermediate Mass Ratio Binaries (E/IMRBs). MSP-E/IMRBs have been identified for their potential for precision parameter estimation in the gravitational strong-field \citep{Wex1999,Liu2012}, whilst also being astrophysically interesting in their own right with regards to theories of stellar formation and evolution. \newline

\noindent The relativistic spin-orbital dynamics of a MSP around a massive BH have been investigated by a variety of authors \citep{Singh2014, Saxton2016,Li2018}. Previous studies have taken GR to be the correct description of reality and so the background spacetime is described by a solution to the Einstein field equations i.e. a Kerr BH. However proof of the `Kerr Hypothesis' - that astrophysical BHs are indeed described by the Kerr solution - is still lacking. An important further consideration is the spin and orbital behaviour when the background spacetime deviates from the Kerr solution, and the corresponding observational signatures. \newline

\noindent  Within GR astrophysical BHs are expected to satisfy the `No Hair Theorem' (NHT), whereby all higher order multipole moments of the gravitational field are expressible as a combination of the two lowest moments, the BH mass ($M$) and spin ($S$). Specifically, should the No Hair Theorem hold then (in geometric units with $c=G=1$) the quadrupole moment $Q$ satisfies,
\begin{eqnarray}
Q = - \frac{S^2}{M}
\end{eqnarray}
Once the mass and spin of the BH are determined, an independent measurement of $Q$ would then provide a direct challenge to the NHT. Whilst its validity would provide another success for GR and rule out alternative theories for which the NHT does not hold, its violation would point to errors in the foundations of relativity, immediately refute the Kerr Hypothesis and may guide the way to alternative theories of gravity. Consequently, being able to accurately determine the observational signature of a non-Kerr quadrupole moment is an essential enterprise. \newline 

\noindent In this work we build on previous investigations to investigate the relativistic orbital spin dynamics of a pulsar around a massive black hole with an arbitrary mass quadrupole. In the extreme mass ratio limit, the MSP mass can be neglected and the orbital and spin dynamics of the MSP are determined by the background spacetime and the interaction of the MSP spin dipole moment with this background metric. In turn, the spacetime is described by the quasi-Kerr metric of \cite{Glampedakis2006}, which describes a stationary, axisymmetric spacetime with a quadrupole moment that deviates slightly from the Kerr value. We determine the effect on the astrophysical observables and discuss the implications for PSR timing in an E/IMRB system and the results for both astrophysics  and fundamental physics.

\section{Equations of motion}
\label{section:MPD}
In this section we construct our framework for describing the orbital dynamics of a spinning MSP around a BH with an arbitrary mass quadrupole. We adopt the natural units, with $c = G = 1$ and normalise the BH mass $M$ such that the gravitational lengthscale $r_g = M = 1$. We use a metric signature $[-,+,+,+]$. Covariant derivatives are denoted by semi-colons e.g. ${T^{\mu \nu}}_{; \mu}$.
\subsection{Quasi-Kerr Metric}
The spacetime of a spinning BH with an arbitrary mass quadrupole moment can be described by the quasi-Kerr metric of \citet{Glampedakis2006}. This metric can be written as, 
\begin{eqnarray}
g_{\mu \nu} = {g^{\rm K}}_{\mu \nu} + \epsilon h_{\mu \nu}
\label{eq:quasiKerr}
\end{eqnarray} 
for Kerr metric ${g^{\rm K}}_{\mu \nu}$, dimensionless deviation parameter $\epsilon$ and the perturbation $h_{\mu \nu}$. To linear order in $\epsilon$, the contravariant form is simply, 
\begin{eqnarray}
g^{\mu \nu} = {g^{\rm K}}^{\mu \nu} - \epsilon h^{\mu \nu}
\end{eqnarray} 
Only the diagonal components of the perturbation are non-zero. The contravariant components in Boyer-Lindquist coordinates are,
\begin{eqnarray}
h^{tt} = \frac{(1-3\cos^2\theta) \mathcal{F}_1(r)}{1-2/r}
\end{eqnarray}
\begin{eqnarray}
h^{rr} = \left (1-\frac{2}{r} \right )(1-3\cos^2\theta) \mathcal{F}_1(r)
\end{eqnarray}
\begin{eqnarray}
h^{\theta \theta} = -\frac{(1-3\cos^2\theta) \mathcal{F}_2(r)}{r^2}
\end{eqnarray}
\begin{eqnarray}
h^{\phi \phi} = -\frac{(1-3\cos^2\theta) \mathcal{F}_2(r)}{r^2 \sin^2\theta}
\end{eqnarray}
where, $\mathcal{F}_{1,2}$ are given in the Appendix of \citet{Glampedakis2006}. The components of the Kerr metric have the usual form,
\begin{eqnarray}
{g^{\rm K}}_{tt} = -\left( 1 - \frac{2r}{\Sigma}\right) 
\end{eqnarray}
\begin{eqnarray}
{g^{\rm K}}_{rr} = \frac{\Sigma}{\Delta}
\end{eqnarray}
\begin{eqnarray}
{g^{\rm K}}_{\theta \theta} = \Sigma
\end{eqnarray}
\begin{eqnarray}
{g^{\rm K}}_{\phi \phi} = \frac{\sin^2 \theta}{\Sigma} [(r^2 + a^2)^2 -\Delta a^2 \sin^2 \theta] 
\end{eqnarray}
\begin{eqnarray}
{g^{\rm K}}_{t \phi} ={g^{\rm K}}_{\phi t} =-\frac{2ar \sin^2 \theta}{\Sigma} 
\end{eqnarray}
where $\Sigma = r^2 + a^2 \cos^2 \theta$ and $\Delta = r^2 - 2r + a^2$ and $a$ is the BH spin parameter ($a = cS/GM^2$). \newline

\noindent The use of this quasi-Kerr metric has a series of advantages. It reduces to the pure Kerr metric in the $\epsilon \rightarrow 0$ limit, it retains axisymmetry and stationarity and is Ricci flat up to quadrupole order (i.e. it is a solution to the Einstein Field equations.) 
%Johansson re regime of validity https://iopscience.iop.org/article/10.1088/0004-637X/716/1/187/pdf

\subsection{MPD Formalism}
In the vacuum Kerr spacetime it is possible via a Hamilton-Jacobi approach to determine the geodesic motion of point particles. However, this is no longer possible for the quasi-Kerr metric, since the perturbed spacetime is no longer Petrov type D \citep{Berti2005} and in general the perturbative quadrupole terms render the Hamiltonian inseparable in the coordinate variables. As a consequence a constant of integration - the Carter Constant \citep{Carter1968} - is lost. Only for the special case of circular equatorial orbits can the Carter constant be recovered. Moreover, even in this special case the geodesic Hamilton-Jacobi approach requires the approximation of the MSP as a spin-less test particle. Such a particle then directly follows a geodesic of the spacetime metric. However real astrophysical objects like pulsars are not point objects, but real bodies with spin. In order to obtain an accurate description of their dynamics, higher-order effects must be considered. In order to account for the extended structure of the MSP and so properly describe the relevant relativistic spin couplings we use the Mathisson-Papatrou-Dixon (MPD) formalism \citep{Mathisson1937,Papapetrou1951,Dixon1974} which has also been used for the modeling of MSP-BH dynamics in a pure Kerr spacetime \citep[e.g.][]{Singh2014,Saxton2016,Li2018}. The spin and orbit of the object are interdependent; by using the MPD approach we can properly account for the spin couplings that arise from a spinning MSP orbiting a spinning BH. \newline

\noindent Starting from the energy-momentum tensor $T^{\mu \nu}$ of the spinning body, the general equation of motion is given by,
\begin{eqnarray}
{T^{\mu \nu}} _{;\nu} = 0 \ ,
\end{eqnarray}
By performing a multipole expansion of the tensor one can construct the `gravitational skeleton' \citep{Dixon1974}. The 0th moment of the expansion is the mass monopole, described by the 4-momentum $p^{\mu}$, whilst the 1st moment is the spin dipole $s^{\mu \nu}$. Since we are operating in the extreme mass ratio limit where the BH mass is much greater than the pulsar mass ($M \gg m$), and the gravitational lengthscale is much grater than the MSP radius ($r_{\rm g} \gg R_{\rm PSR}$) then the MSP spin dynamics can be described by the background gravitational field and the dynamical spin interaction with this field. Moreover, moments of the multipole expansion greater than the quadrupole can be neglected, since the motion is dominated by the lower order terms \citep{Singh2014}. In this case the corresponding differential equations for the evolution of the first two moments are \citep{Mathisson1937,Papapetrou1951, Dixon1974},
\begin{eqnarray}
 \frac{D p^{\mu}}{d \tau}= - \frac{1}{2} {R^{\mu}}_{\nu \alpha \beta} u^{\nu} s^{\alpha \beta} \ ,
\label{Eq:mpd1}
\end{eqnarray}
\begin{eqnarray}
 \frac{D s^{\mu \nu}}{d \tau}= p^{\mu} u^{\nu} -p^{\nu} u^{\mu} \ ,
\label{Eq:mpd2}
\end{eqnarray}
where $\tau$ is the proper time along the MSP worldline, $u^{\nu}$ is the PSR 4-velocity and ${R^{\mu}}_{\nu \alpha \beta}$ the Riemann curvature tensor. In order to close this system of equations it is necessary to specify a spin supplementary condition (SSC). This is due to the fact that when performing the multipole expansion of the energy-momentum tensor, one must specify a worldline about which to do the expansion. Whilst this worldline would usually map to the centre of mass of the body, in GR the centre of mass of a spinning body is not invariant. The choice of SSC is therefore equivalent to choosing an observer with respect to which the centroid is defined. Different SSC choices are possible, and for our purposes we adopt the Tulczyjew-Dixon (TD) condition, 
\begin{eqnarray}
s^{\mu \nu} p_{\nu} = 0 \ ,
\label{Eq:ssc}
\end{eqnarray}
\citep{Tulczyjew1959,Dixon1964}. This is equivalent to choosing the centre of mass as measured in the zero 3 momentum frame. This SSC is advantageous since it specifies a unique worldline, whilst other choices of SSC are infinitely degenerate \citep[for discussion see][]{Costa2014}. With this SSC, the mass of the MSP
\begin{eqnarray}
m = \sqrt{-p^{\mu} p_{\mu}}
\end{eqnarray}
is also conserved. \newline 

\noindent The Moller radius $R_{\rm Moller}$ describes the radius of the disk which contains the set of all the potential positions of the centre of mass as measured by an observer,
\begin{eqnarray}
R_{\rm Moller} = \frac{s}{m}
\end{eqnarray}
where $s^2 =  s^{\mu \nu } s_{\mu \nu} / 2$ is a conserved scalar quantity. Since $R_{\rm Moller} \ll R_{\rm PSR}$, it follows that  the pole-dipole terms are much stronger than the dipole-dipole terms \citep{Singh2014}. Therefore, to first order the 4-velocity and 4 momentum are parallel, i.e. $p^{\mu} \approx m u^{\mu}$ and the equations of motion become,
\begin{eqnarray}
{u^{\mu}}_{; \tau}= - \frac{1}{2m} {R^{\mu}}_{\nu \alpha \beta} u^{\nu} s^{\alpha \beta} \ ,
\end{eqnarray}
\begin{eqnarray}
{s^{\mu \nu}}_{;\tau}\approx 0 \ ,
\end{eqnarray} 
\citep{Chicone2005,Mashhoon2006}. 
The ordinary differential equations to then be integrated are \citep{Singh2005,Mashhoon2006}:
\begin{eqnarray}
\frac{dp^{\alpha}}{d\tau} = - \Gamma_{\mu\nu}^{\alpha} p^{\mu}u^{\nu} + \lambda \left( \frac{1}{2m} R^{\alpha}_{\beta \rho \sigma} \epsilon^{\rho \sigma}_{\quad \mu \nu} s^{\mu} p^{\nu} u^{\beta}\right) \ ,
\label{eq:ode1}
\end{eqnarray}

\begin{eqnarray}
\frac{ds^{\alpha}}{d \tau} = - \Gamma^{\alpha}_{\mu \nu} s^{\mu}u^{\nu} + \lambda \left(\frac{1}{2m^3}R_{\gamma \beta \rho \sigma} \epsilon^{\rho \sigma}_{\quad \mu \nu} s^{\mu} p^{\nu} s^{\gamma} u^{\beta}\right)p^{\alpha} \ ,
\label{eq:odeS}
\end{eqnarray}

\begin{eqnarray}
\frac{dx^{\alpha}}{d\tau} = -\frac{p^{\delta}u_{\delta}}{m^2} \left[ p^{\alpha} + \frac{1}{2} \frac{\lambda (s^{\alpha \beta} R_{\beta \gamma \mu \nu} p^{\gamma} s^{\mu \nu})}{m^2 + \lambda(R_{\mu \nu \rho \sigma} s^{\mu \nu} s^{\beta \sigma}/4)}\right] \ ,
\label{eq:ode2}
\end{eqnarray}
where $s^{\mu}$ is the spin 4-vector given by,
\begin{eqnarray}
s_{\mu} = \frac{1}{2m} \epsilon_{ \mu \nu \alpha \beta} p^{\nu} s^{\alpha \beta}
\label{eq:spinvector}
\end{eqnarray}
and the dimensionless parameter $\lambda$ is used to label the terms which contribute to MPD spin-curvature coupling ( i.e. $\lambda = 1$ includes spin-curvature coupling, for $\lambda = 0$ the coupling is omitted). In the $\lambda \rightarrow 0$ limit the conventional spin-spin and spin-orbit couplings are recovered.

\subsection{Initial Conditions}
\noindent The initialization of $p^{\mu}$ is tantamount to specifying the sort of orbit that we want to describe. In a pure Kerr spacetime is is possible to map the Keplerian orbital elements $\Theta = (\mathcal{P},e,i)$ (semi-latus rectum, eccentricity and inclination respectively) to the conserved quantities $E, L, Q$ (energy, angular momentum, Carter constant) \citep{Schmidt2002}. In turn, these constants of the motion can be related to the orbital 4-momentum. However, the quasi-Kerr metric is not in general separable and so it is not possible to proceed in the same way. In order to enforce separability and map between the sorts of orbits we want to describe (i.e. specify $\mathcal{P},e,i$) and the $p^{\mu}$ initialization we restrict our analysis to the equatorial plane ($i=0$). In this case, we can write,
\begin{align*}
(u^r)^2 = V(r) = \left[(r^2 + a^2)E -aL\right]^2 - \Delta \left[r^2 + (L - aE)\right] \\
-\epsilon r^4 \left(1 - \frac{2}{r}\right)\left((\mathcal{F}_3 - \mathcal{H}_3)\frac{L^2}{r^2} + \mathcal{F}_3\right)
\end{align*} 
where $\mathcal{F}_3$ and $\mathcal{H}_3$ are defined in the Appendix of \citep{Glampedakis2006}. Collecting terms is is evident that we can write,
\begin{eqnarray}
V(r) = f(r) E^2 - 2g(r) EL - h(r) L^2 -d(r)
\label{eq:schmidt}
\end{eqnarray}
where
\begin{eqnarray}
f(r) = r^4 + a^2 r (r+2)
\end{eqnarray}
\begin{eqnarray}
g(r) = 2ar
\end{eqnarray}
\begin{eqnarray}
h(r) = r(r-2) -\epsilon\left[2 \mathcal{F}_3 r - 2\mathcal{H}_3r -\mathcal{F}_3 r^2 +\mathcal{H}_3 r^2\right]
\end{eqnarray}
\begin{eqnarray}
d(r) = r^2 \Delta - \epsilon \left[2 \mathcal{F}_3 r^3 - \mathcal{F}_3 r^4 \right]
\end{eqnarray}
Eq \ref{eq:schmidt} is of he same form as the base construction in \cite{Schmidt2002} and so we can proceed in an analagous way and so define the energy and angular momentum as,
\begin{eqnarray}
E=\sqrt{\frac{\kappa \rho+2 \epsilon \sigma-2 D \sqrt{\sigma\left(\sigma \epsilon^{2}+\rho \epsilon \kappa-\eta \kappa^{2}\right)}}{\rho^{2}+4 \eta \sigma}}
\end{eqnarray}
\begin{eqnarray}
\tilde{L}_{z}=-\frac{g_{1} E}{h_{1}}+\frac{D}{h_{1}} \sqrt{g_{1}^{2} E^{2}+\left(f_{1} \tilde{E}^{2}-d_{1}\right) h_{1}}
\end{eqnarray}
where
\begin{eqnarray} 
\kappa & \equiv d_{1} h_{2}-d_{2} h_{1} 
\end{eqnarray}
\begin{eqnarray}
\varepsilon  \equiv d_{1} g_{2}-d_{2} g_{1} 
\end{eqnarray}
\begin{eqnarray}
\rho  \equiv f_{1} h_{2}-f_{2} h_{1} 
\end{eqnarray}
\begin{eqnarray}
\eta  \equiv f_{1} g_{2}-f_{2} g_{1} 
\end{eqnarray}
\begin{eqnarray}
\sigma & \equiv g_{1} h_{2}-g_{2} h_{1} 
\end{eqnarray}
and 
\begin{eqnarray}
(f_n, g_n, h_n, d_n) \equiv (f(r_n), g(r_n), h(r_n), d(r_n))
\end{eqnarray}
with $D = \pm 1$ to denote prograde and retrograde orbits and $r_1$ denotes the periapsis radius and $r_2$ the apoapsis. With the orbital constants specified, the initialization of $p^{\mu}$ is given by $p^{\mu} = m u^{\mu}$ where \cite{Glampedakis2006},
\begin{eqnarray}
u^t = \frac{1}{r^2}\left((r^2 + a^2) \frac{\mathcal{P}}{\Delta}   - a(aE - L) \right) - \epsilon \left(1 - \frac{2}{r}\right)^{-1} \mathcal{F}_3 E
\end{eqnarray}
\begin{align*}
(u^r)^2= &\frac{1}{r^4} \Big ( \mathcal{P}^2 - \Delta(r^2+ (L-aE)^2) \\
&-\epsilon r^4 \left(1 -\frac{2}{r}\right) ((\mathcal{F}_3 - \mathcal{H}_3)\frac{L^2}{r^2}+ \mathcal{F}_3) \Big )
\end{align*}
\begin{eqnarray}
u^{\theta} = 0
\end{eqnarray}
\begin{eqnarray}
u^{\phi} = \frac{1}{r^2} \left(\frac{a \mathcal{P}}{\Delta} -aE +L\right) -\epsilon  \frac{\mathcal{H}_3 L}{r^2}
\end{eqnarray}
This framework of mapping the geometric orbital parameters to the initial conditions on $p^\mu$ is fundamentally an approximation since it does
not include spin effects of the MSP. As a consequence the orbital parameters are not constant as they would be for a weak-field Keplerian orbit, but vary in time \citep[see e.g.][]{Singh2014}. Nevertheless, these variations are typically small and so this mapping framework provides a decent first-order approximation to the sorts of orbits that we want to model. \textbf{It is also worth noting that even in the case where the underlying metric is separable \citep[e.g. ][]{Johannsen2013} chaotic motion can occur should the orbital motion depart from being equatorial. }\newline

\section{Spin-Orbit Dynamics of a MSP}
We take as our canonical MSP a NS with mass $1.4 M_{\odot}$ and spin period $1$ ms. We will consider BHs of both `intermediate' mass $( M = 10^3 M_{\odot})$ and `supermassive' $( M = 10^6 M_{\odot})$. Since the spin of astrophysical BHs is not well observationally constrained we set the BH spin at an intermediate value of $a= \pm 0.6$. We will consider different orbital parameters, $\Theta = (\mathcal{P}, e)$, and explore the effects induced by the quadrupole moment and the spin couplings. We define $\delta_{\epsilon} X$ as the quadrupole-induced difference on a general quantity $X$ and similarly $\delta_{\lambda} X$ as the difference the quantity induced by the MSP spin, i.e.
\begin{eqnarray}
\delta_{\epsilon} X = X(\epsilon,\Theta) - X(0,\Theta)
\end{eqnarray}
\begin{eqnarray}
\delta_{\lambda} X = X(\lambda,\Theta) - X(0,\Theta)
\end{eqnarray}

\subsection{Orbital Dynamics}
\begin{figure}
	\includegraphics[width=\columnwidth]{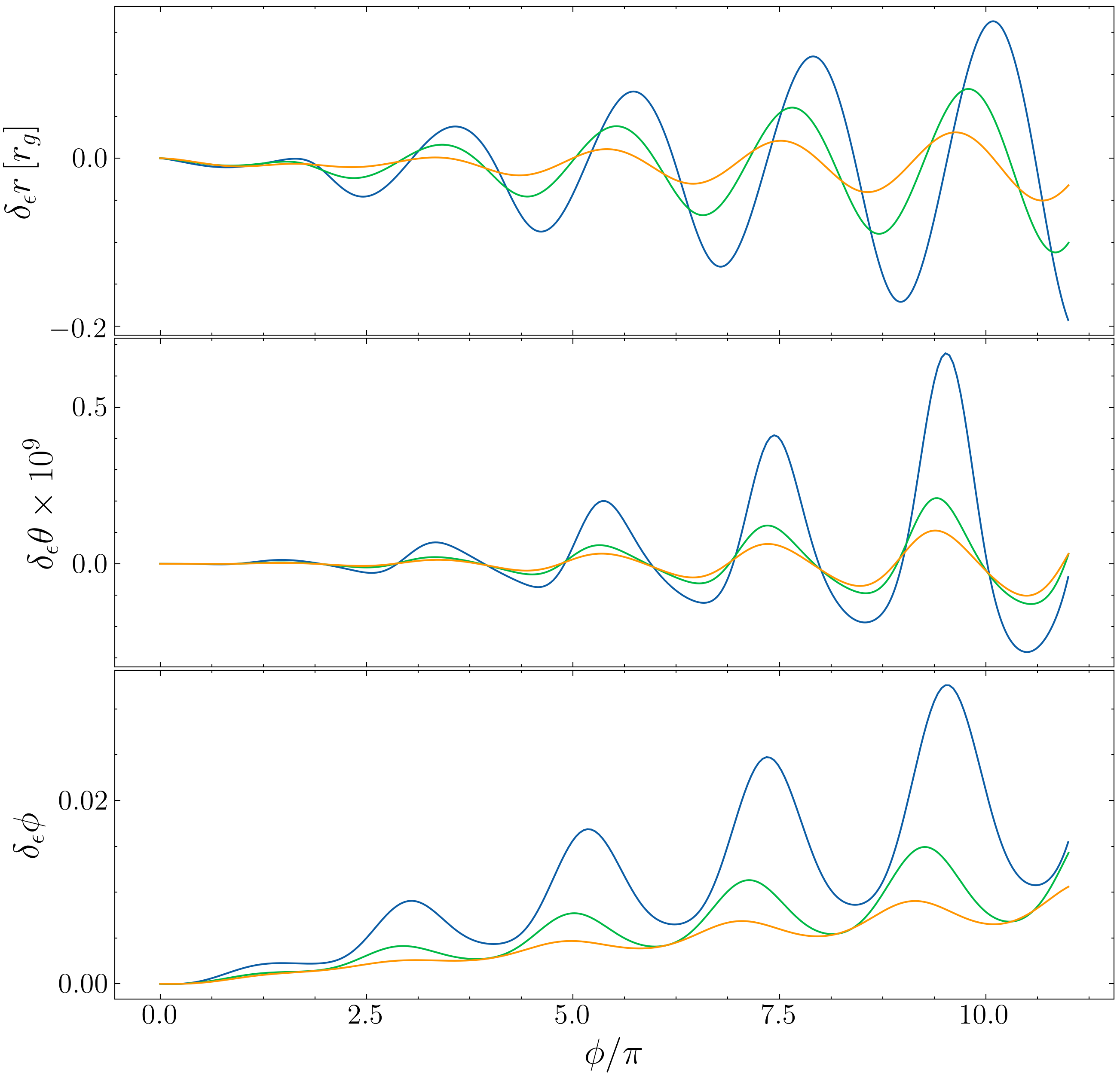}
	\caption{Variation induced in the coordinate variables due to a non-Kerr quadrupole moment $\epsilon = 0.1$ with respect to the Kerr solution ($\delta x = x_{\epsilon=0.1} - x_{\epsilon = 0.0}$). The BH has $M=4.3 \times 10^6 M_{\odot}$, $a=0.6$ and the MSP has orbital parameters $a_*=50$, $ i = 0$ and eccentricities $e=0.2,0.4,0.6$ (orange, green, blue respectively). More eccentric orbits exhibit greater magnitude deviations induced by $\epsilon$ on account of their closer periapsis passage. The $x$-axis is the $\phi$ coordinate when $\epsilon = 0.1$.}
	\label{fig:a}
\end{figure}

The BH quadrupole moment modifies the background spacetime, whilst the MSP spin couplings alter the interaction of the MSP spin with this background field. As a consequence the orbital trajectory of the MSP exhibits different behaviour compared to the pure Kerr geodesic case. The influence of a non-Kerr quadrupole moment ($\epsilon = 0.1$) on the spatial coordinate variables is presented in Figure \ref{fig:a} for a system with a supermassive BH with spin parameter $a=0.6$ orbited by a MSP in the equatorial plane with semi-major axis $a_*=50 \, r_{\rm g}$ at different eccentricities.
The influence of the quadrupole has two clear types of modification on the orbital trajectory, which are most apparent by examining the behaviour of the $\phi$ coordinate variable. The first is a longer timescale secular drift, which causes a precession of the orbital orientation (the angle of pericentre). In addition to this secular behaviour, there is also a short timescale, periodic feature as the MSP passes through periapsis. These periodic features are strongest for more eccentric orbits, since these systems have shorter periapsis passages. There is a large magnitude in the quadrupole-induced variation in the $r$ and $\phi$ coordinate variables, with a fractional difference of order $1 \%$. In absolute terms, these spatial differences are of the order $0.1 r_{\rm g}$. The $\delta_{\epsilon} r$ and $\delta_{\epsilon} \phi$ evolution is out of phase by $\pi/4$; the maxima of  $\delta_{\epsilon} \phi$ occur as the MSP passes through periapsis whilst the maxima of $\delta_{\epsilon} r$ occurs when $\delta_{\epsilon} \phi$ is changing most rapidly. \newline

\begin{figure}
	\includegraphics[width=\columnwidth]{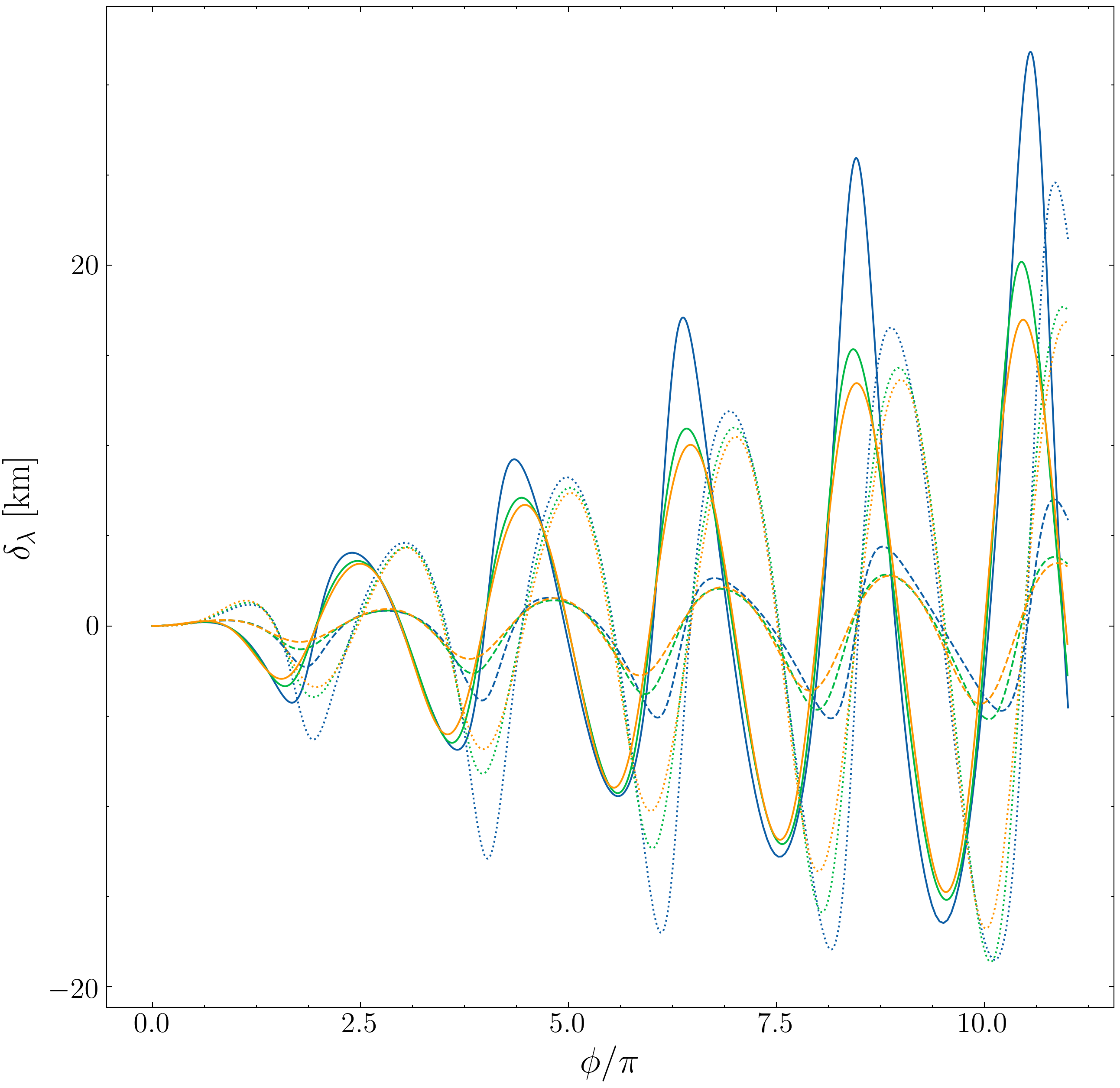}
	\caption{Difference in the $x,y,z$ coordinate variables (solid, dotted, dashed lines respectively) between a fast-spinning and non-spinning MSP, on a background quasi-Kerr spacetime with $\epsilon = 0.1$. The eccentricities are $e=0.2,0.4.0.6$ (orange,green, blue respectively), with orbital parameters $a_*=50 r_{\rm g}$, $i = 0$ whilst the BH mass $M=4.3 \times 10^6 M_{\odot}$, $a=0.6$. Periodic variations are seen as the MSP passes through periapsis, with greater magnitude variations for more eccentric orbits.}
	\label{fig:b}
\end{figure}

\noindent For the same system, we can also examine the spin-induced variation in the MSP coordinate position working in Cartesian coordinates, rather than spherical-polar Boyer Lindquist coordinates (Figure \ref{fig:b}). The two coordinate systems are mapped via the standard spherical polar relation as,
\begin{eqnarray}
x = m \sin\theta \cos \phi
\end{eqnarray}
\begin{eqnarray}
y = m \sin\theta \sin \phi
\end{eqnarray}
\begin{eqnarray}
z = r \cos \theta 
\end{eqnarray}
for $m = \sqrt{r^2 +a^2}$. Again we observe short timescale periodic features as the MSP passes through periapsis, with the effect being stronger for more eccentric orbits.  The spin couplings cause a variation not just in the vertical coordinate $\hat{z}$ direction \citep[e.g.][]{Singh2014}, but also variations in the in-plane $x-y$ motion. Across the considered parameter space, typical variations are of magnitude $\mathcal{O}(10)$ km. Whilst this lengthscale is small compared to the gravitational lengthscale $r_g$, a $10$ km variation is equivalent to a light travel time of $\sim 30 \mu s$ which is readily detectable via radio pulsar timing. The astrophysical implications of these spin couplings will be further explored later. Whilst we have just considered here a BH with mass $\sim 10^6 M_{\odot}$, the magnitude of the spin-curvature coupling is independent of  the BH mass. \newline 

\begin{figure}
	\includegraphics[width=\columnwidth]{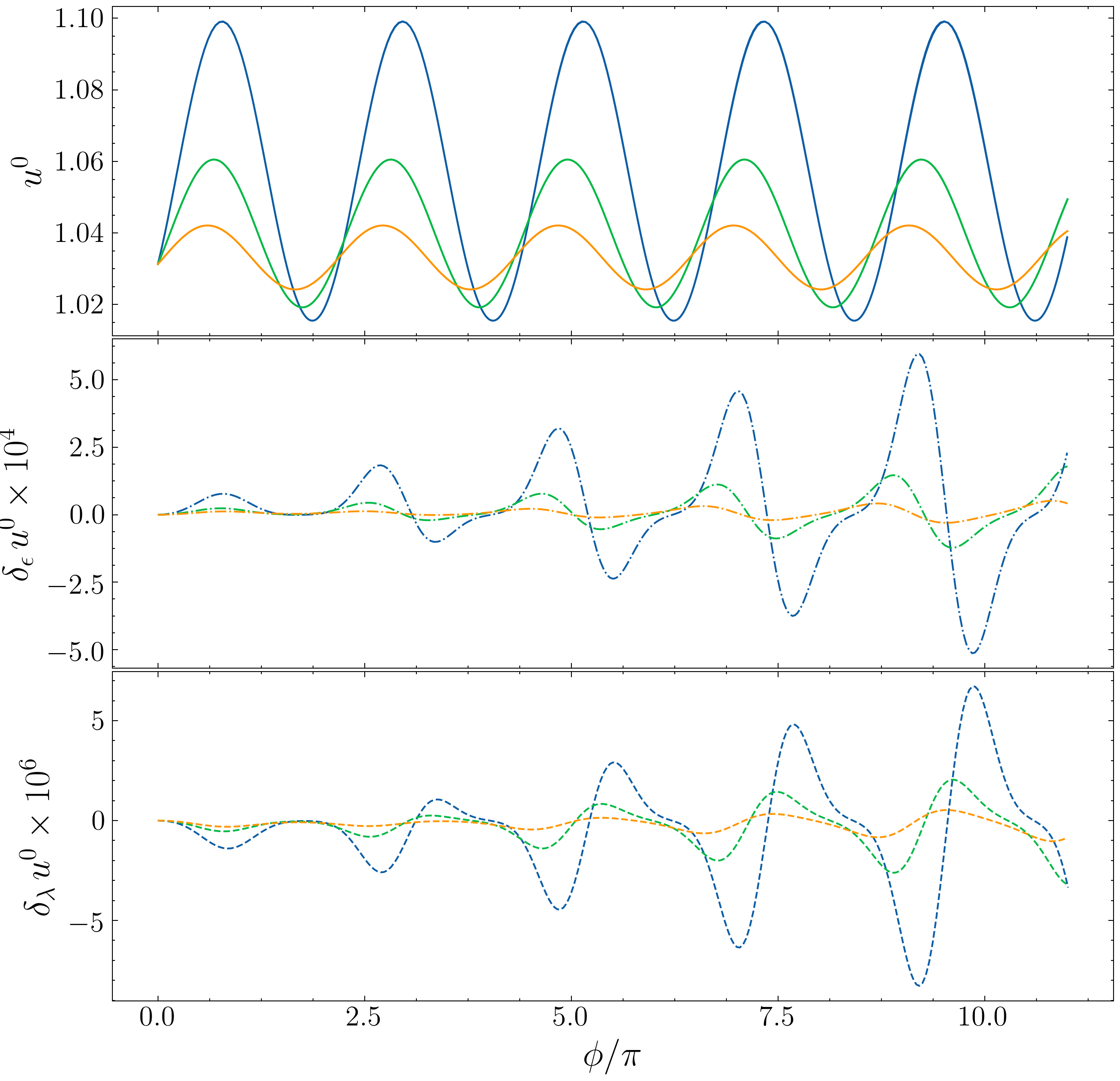}
	\caption{The 0-th component of the MSP 4-velocity (top panel) for an MSP with orbital parameters $r=50$, $\iota = 0$, $e=0.2,0.4.0.6$ (blue,orange,green respectively) around a BH of mass $M= 10^3 M_{\odot}$ and $a=0.6$. Middle panel shows the corrections due to a non-Kerr quadrupole moment of $\epsilon=0.1$ and the bottom panel shows the corrections due to the MSP spin couplings on a background quasi-Kerr spacetime.}
	\label{fig:c}
\end{figure}

\noindent In addition to the spatial evolution of the MSP, we can also consider the time evolution i.e. the ratio of the proper time $\tau$ to the coordinate time $t$. The nature of MSPs as relativistic precision clocks means that this difference can be directly measured, and indeed the difference in the rate at which the MSP `ticks' due dynamical redshift is a key component of pulsar timing models (the `Einstein delay'). This ratio is given by the 0-th component of the MSP 4-velocity ($u^0 = dt/d\tau$). The evolution of $u^0$ is shown in Figure \ref{fig:c}, for a BH with $M = 10^3 M_{\odot}$, $a=0.6$ along with the quadrupole and spin ($\epsilon - \lambda$) induced corrections.
The $\epsilon-\lambda$ induced variations in $u^0$ exhibit the same general behaviour as the variations induced in the spatial coordinates; $\delta u^0$ displays periodic oscillations, with the frequency set by the orbital frequency of the system (extrema of $\delta u^0$ at periapsis), and more eccentric orbits displaying greater magnitude variations. The corrections due to the quadrupole moment, $\mathcal{O} (10^{-4})$, are greater than those due to the spin couplings, $\mathcal{O}(10^{-6})$. 

\subsection{Spin Dynamics}

Since generally the MSP spin axis is not aligned with the orbital angular momentum axis, the spin vector $s^{\mu}$ evolves with time and the spin axis exhibits precession and nutation. This spin evolution in turn influences the MSP orbital dynamics (e.g. Eq. \ref{eq:ode1} - Eq. \ref{eq:ode2}). In a Newtonian description, the spin 3-vector $\boldsymbol{s}_1$ of an object of mass $m_1$ in a binary systems with another object of mass $m_2$, spin vector $\boldsymbol{s}_2$ is given by \citep{Kidder1995},
\begin{eqnarray}
\dot{\boldsymbol{s}}_{1}= \frac{1}{r^3} \left[ (\boldsymbol{L} \times \boldsymbol{s}_1) \left(2 + \frac{3}{2}\frac{m_1}{m_2}\right) - \boldsymbol{s}_1\times \boldsymbol{s}_2 + 3\left(\hat{\boldsymbol{n}} \cdot \boldsymbol{s}_2 \right) \left(\hat{\boldsymbol{n}} \times \boldsymbol{s}_1 \right) \right]
\end{eqnarray}
where $\hat{\boldsymbol{n}}$ is the unit vector between the two bodies and $\boldsymbol{L}$ is the usual orbital angular momentum. The first term $(\boldsymbol{L} \times \mathbf{s}_1)$ describes the  spin-orbit coupling and the other terms describe the spin-spin couplings. In a general relativistic context, the geodetic precession velocity of a gyroscope can be generally represented as \citep[e.g. ][]{Connell1969},
\begin{eqnarray}
\Omega = \Omega_{\rm DS} + \Omega_{\rm LT} + \Omega_{\rm Q}
\label{eq:geoprec}
\end{eqnarray}
where $\Omega_{\rm DS}$, $\Omega_{\rm LT}$, $\Omega_{\rm Q}$ are the de Sitter (precession due to the BH mass), Lense-Thirring (due to the BH spin) and quadrupole contributions respectively. Via the MPD formalism on our quasi-Kerr spacetime we can consistently describe both the geodesic and spin coupling effects simultaneously. \newline 

\begin{figure}
	\includegraphics[width=\columnwidth]{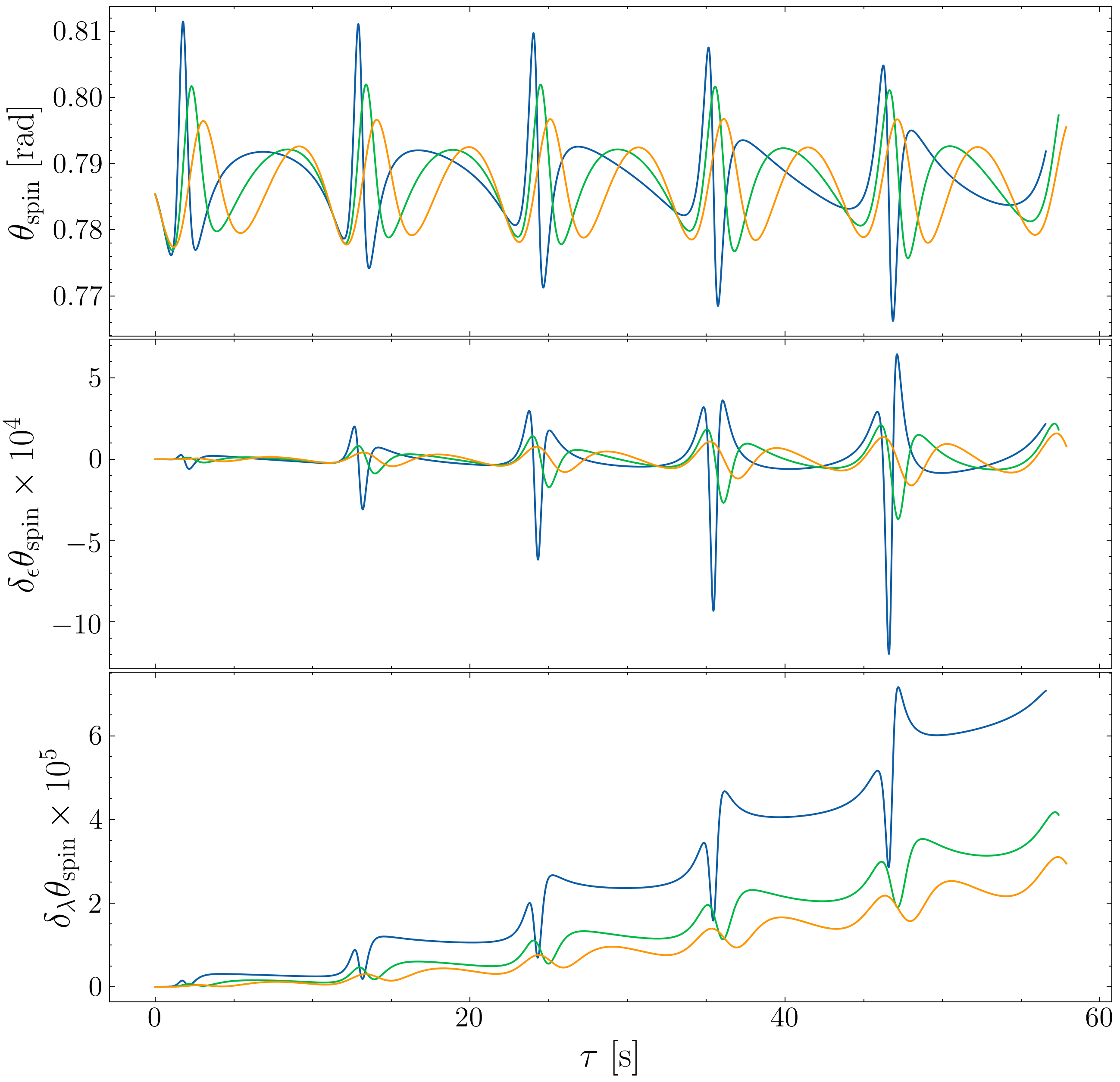}
	\caption{\textit{(Top panel:)} The nutation of the MSP spin axis due to geodetic and spin coupling effects for a MSP with spin period $1$ ms, semi-major axis $50 r_{\rm g}$ around a BH with $M = 10^3 M_{\odot}, a = +0.6$, and eccentricities $e=(0.6,0.4,0.2)$ (blue green,orange). Rapid variations are seen as the MSP passes through periapsis, with larger magnitude oscillations for more eccentric orbits. \textit{Middle panel:} Quadrupole -induced corrections to the nutation. These are periodic and maximal when the MSP is closest to the BH \textit{Bottom panel:} Spin-induced corrections to the nutation. These again are periodic and also display a secular behaviour.}
	\label{fig:spinTHETA}
\end{figure}

\begin{figure}
	\includegraphics[width=\columnwidth]{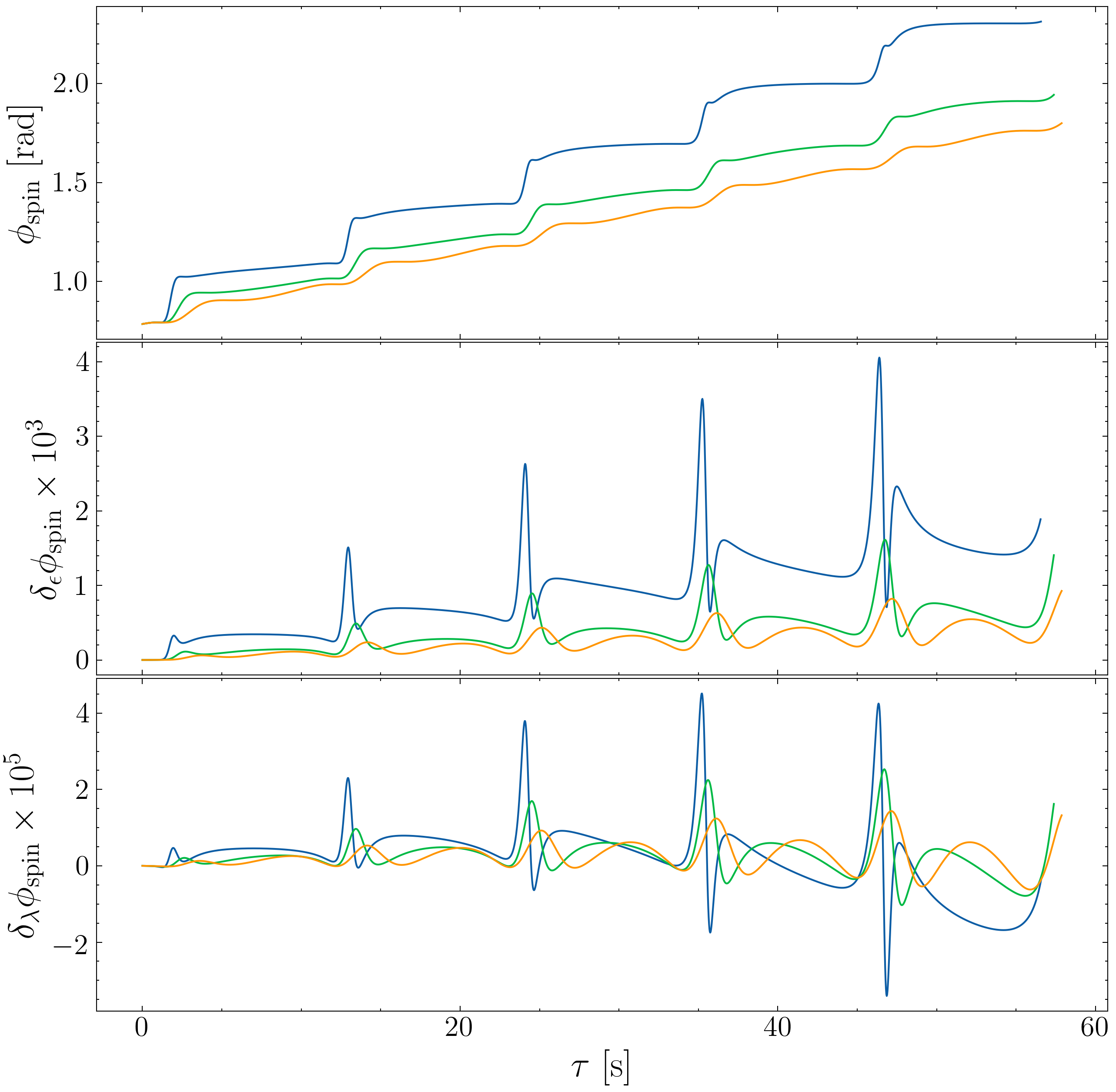}
	\caption{As Fig. \ref{fig:spinTHETA}, but for the precession of the MSP spin axis. The precession exhibits rapid `jumps' as the MSP goes through periapsis which causes a general secular, step-wise increase in the precession angle. The corrections due to $\epsilon - \lambda$ effects exhibit the same general behaviour as in the nutation, }
	\label{fig:spinPHI}
\end{figure}

\noindent The orientation of the spin axis can be described by the two Euler angles, $\theta_{\rm spin}$, which relates to the nutation and $\phi_{\rm spin}$ which describes the precession. These angles are given in the laboratory frame as,
\begin{eqnarray}
\theta_{\rm spin}  =\atantwo \left( \sqrt{S_x^2 + S_y^2}, S_z\right)
\end{eqnarray}
\begin{eqnarray}
\phi_{\rm spin}= \atantwo \left(S_y, S_x\right)
\end{eqnarray}
where $S_{x,y,z}$ are the Cartesian components of the spin vector, related to $s_{\mu}$ as (see e.g. Eq.\ref{eq:odeS}. Eq. \ref{eq:spinvector}),
\begin{eqnarray}
S_x = s_1 \sin (\theta) \cos(\phi) +s_2 r \cos(\theta) \cos(\phi) -s_3 r  \sin(\theta) \sin(\phi)
\end{eqnarray}
\begin{eqnarray}
S_y = s_1 \sin (\theta) \sin(\phi) +s_2 r \cos(\theta) \sin(\phi) +s_3 r  \sin(\theta) \cos(\phi)
\end{eqnarray}
\begin{eqnarray}
S_z = s_1 \cos(\theta) -s_2 r \sin(\theta) 
\end{eqnarray}
 The spin evolution in time of an MSP orbiting (equatorial, semi-major axis = $50 r_{\rm g}$) an IMBH  ($M = 10^3 M_{\odot} \, , \,  a=+0.6$) are illustrated in Figs. \ref{fig:spinTHETA}, \ref{fig:spinPHI}, over 5 orbits, along with corrections induced by spin couplings and the BH quadrupole moment, where initially $\theta_{\rm spin} = \phi_{\rm spin} = \pi/4$. For an MSP on a Kerr geodesic, $\theta_{\rm spin}$ is described by Eq. \ref{eq:geoprec} i.e. the nutation is governed solely by geodesic effects determined by the background spacetime. In this case $\theta_{\rm spin}$ exhibits rapid,periodic variations as the MSP passes through periapsis. More eccentric orbits display greater magnitude variations in $\theta_{\rm spin}$, whilst also being more constrained in time. The introduction of a BH quadrupole with $\epsilon = 0.1$ induces additional contributions, with $\delta_{\epsilon} \theta_{\rm spin} \sim \mathcal{O} (10^{-4})$. These contributions are periodic with the periodicity set by the MSP orbital frequency. Since the strength of the quadrupole interactions is governed by the distance of the MSP from the central BH, more eccentric orbits (with closer periapsis distances) exhibit the largest magnitude variations. These large magnitude oscillations decay more rapidly than oscillations from less eccentric orbits and as a consequence $\delta_{\epsilon} \theta_{\rm spin}$ can be greater at certain orbital phases for less eccentric systems. The spin-induced variations in $\theta_{\rm spin}$ follow the same general pattern with rapid, periodic perturbations as the MSP passes through periapsis, albeit with a different time profile. In addition the $\lambda$-perturbations display a secular, long timescale behaviour due to the drastic change in $\theta_{\rm spin}$ as the MSP comes out of periapsis. This is unlike the $\epsilon$ perturbations whereby $\delta_{\epsilon} \theta_{\rm spin} \rightarrow \sim 0$ after the MSP has gone through periapsis. These spin-induced variations for the systems considered here are typically an order of magnitude smaller than those due to the quadrupole moment. 

The precession of the spin axis $\phi_{\rm spin} (\tau)$, follows the same general behaviour as $\theta_{\rm spin}$ with rapid, large amplitude, periodic variations as the MSP goes through periapsis. However, there is also an additional secular contribution that causes $\phi_{\rm spin}$ to generally increase with time. Moreover, the rapid change in $\phi_{\rm spin}$ at periapsis is not oscillatory in the same way that $\theta_{\rm spin}$ is, but can instead be seen to be a rapid `jump'. The $\epsilon-\lambda$ induced variations also display the same general behaviour as in the $\theta_{\rm spin}$ case, with $\delta_{\epsilon} \phi_{\rm spin} \sim \mathcal{O} (10^{-3})$ and $\delta_{\lambda} \phi_{\rm spin} \sim \mathcal{O} (10^{-5})$. However, this time the $\epsilon$-corrections exhibit an additional secular contribution, whilst the $\lambda$-corrections are periodic.

\begin{figure}
	\includegraphics[width=\columnwidth]{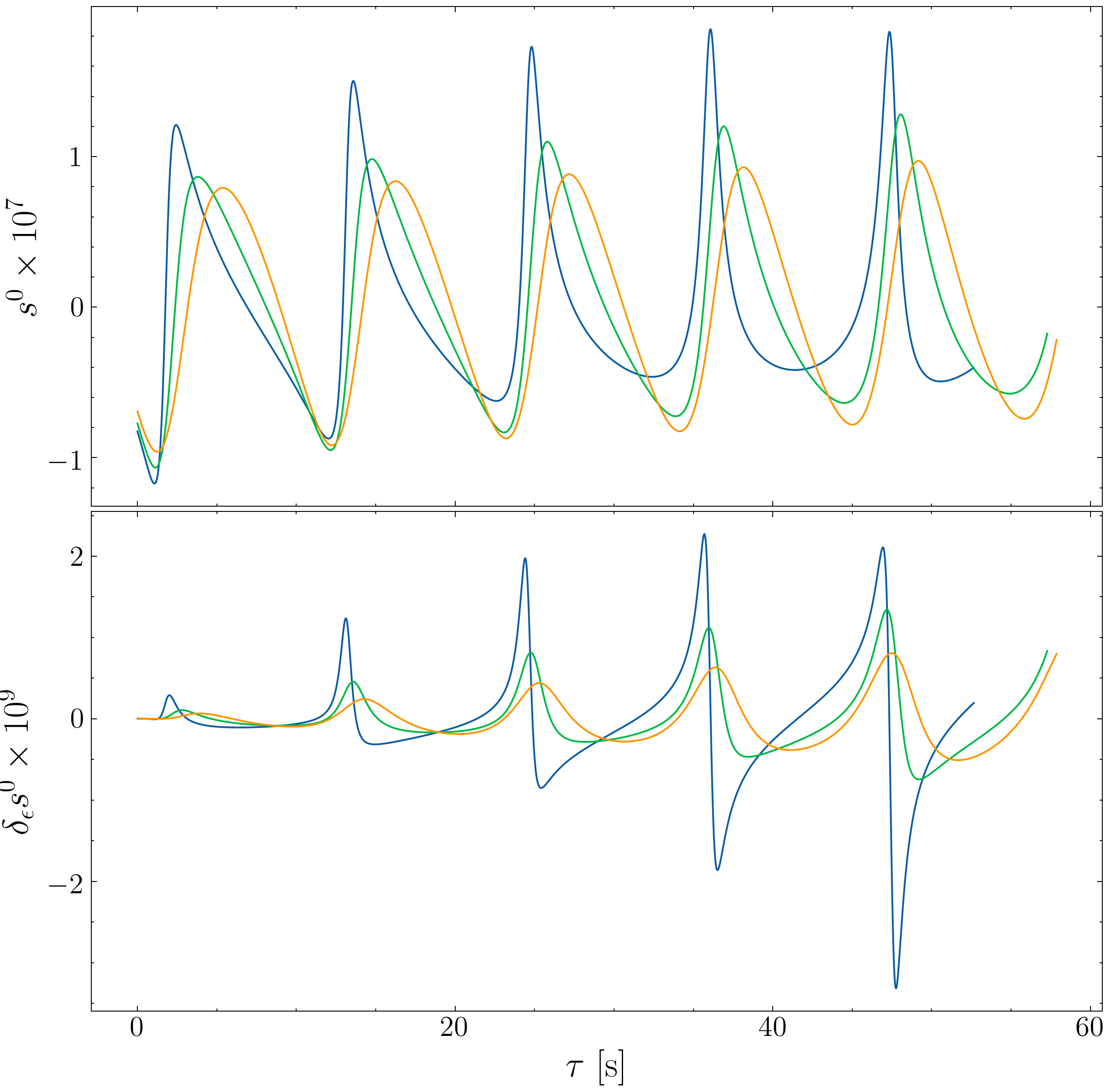}
	\caption{\textit{Top panel:} Time evolution of $s^0$ due to the breaking of the spacetime rotational symmetry as a consequence of the MSP spin. Rapid variations in $s^0$ are seen as the MSP passes through periapsis, with greater amplitude oscillations for more eccentric systems. \textit{Bottom panel:} The difference in $s^0$ induced by a quadrupole moment $\epsilon = 0.1$ as compared to the Kerr $(\epsilon = 0)$ case. The quadrupole moment induces a relative error on the order of $1 \%$. The BH has parameter $M = 10^3 M_{\odot}, a = -0.6$.}
	\label{fig:spinS0}
\end{figure}

For the same system in a retrograde orbit ($a = -0.6$), the evolution of the time component of the spin vector, $s^0$, is displayed in Fig. \ref{fig:spinS0}, along with the variations induced by the quadrupole moment of the BH. This variation in $s^0$ is a strong-field relativistic phenomenon, since this component would remain constant in the usual PN formulation \citep[see discussion of this issue in ][]{Li2018}. The exact physical meaning of the temporal component  of the spin vector $s_0$ is not well understood, although it can be shown to have some relation to the difference in the centre of mass and the centre of momentum. In addition, from the SSC (Eq. \ref{Eq:ssc}) it can be shown by dividing by $u^0$ that the time component of the covariant form of the spin vector is
\begin{eqnarray}
s_0 = -\left(s_1 \frac{dr}{dt}+ s_2 \frac{d \theta}{dt}+ s_3 \frac{d \phi}{dt}\right)
\end{eqnarray}
which describes the spatial components of a spin vector (i.e. the 3-vector) as measured by a static observer.Since $u^0$ is related to the relativistic time dilation, $s_0$ may also be related to the relativistic aberration of light. From Fig. \ref{fig:spinS0} we can see that for each of the eccentric orbits $s^0$ oscillates with rapid variations as the MSP goes through periapsis. The presence of the quadrupole moment induces an additional variation in $s^0$ on the scale of $\sim 1 \%$.

\section{Astrophysical/Observational Implications}

\subsection{Radio Pulsar Timing}
There are expected to be large populations of MSPs at the centre of the Galaxy \citep{Wharton2012,Rajwade2017} whilst Globular clusters are also known to harbour large numbers of MSPs \citep{Hui2010,Pan2016}. The detection of a MSP with a BH companion in the centre of these stellar clusters is a major target for advanced radio facilities such as the SKA \citep{Combes2015} or the NASA Deep Space Network \citep{Majid2019,Pearlman2019}. Whilst these systems are scientifically rich, they also inhabit the gravitational strong field and so radio timing in these regimes encounters additional challenges not experienced by standard weak-field pulsar observations \citep[see e.g. ][]{Kimpson2019B}. We will now discuss the implications of the spin-orbital dynamics for strong-field pulsar astronomy.
\subsubsection{Implications of Orbital Dynamics}
The orbital motion of the MSP is determined by the background spacetime, and the dynamical spin interaction of the MSP with this gravitational field. Consequently, as we have shown, the quadrupole moment of the central massive BH and the spin couplings of the MSP will lead to variations in the coordinate position of the MSP. In turn, the variation in the coordinate variables of the MSP (e.g. Fig. \ref{fig:a}) compared to the geodesic case will manifest observationally in the radio MSP timing solution. Firstly the additional $\epsilon - \lambda$ contributions will cause pericentre angle $\omega$ and the projected semi-major axis $x = a \sin i / c$ to exhibit a secular evolution \citep[see e.g.][]{Wex1999}. The magnitude of this effect is sub-dominant to the effect of the BH mass and spin and so as noted in \cite{Wex1999} these secular effects may not be a useful may to actually measure the BH quadrupole. However, when constructing a complete MSP timing solution over longer time scales and several orbital periods it will be import to include the contributions of the spin couplings and the BH quadrupole to the orbital precession rate.

\begin{figure}
	\includegraphics[width=\columnwidth]{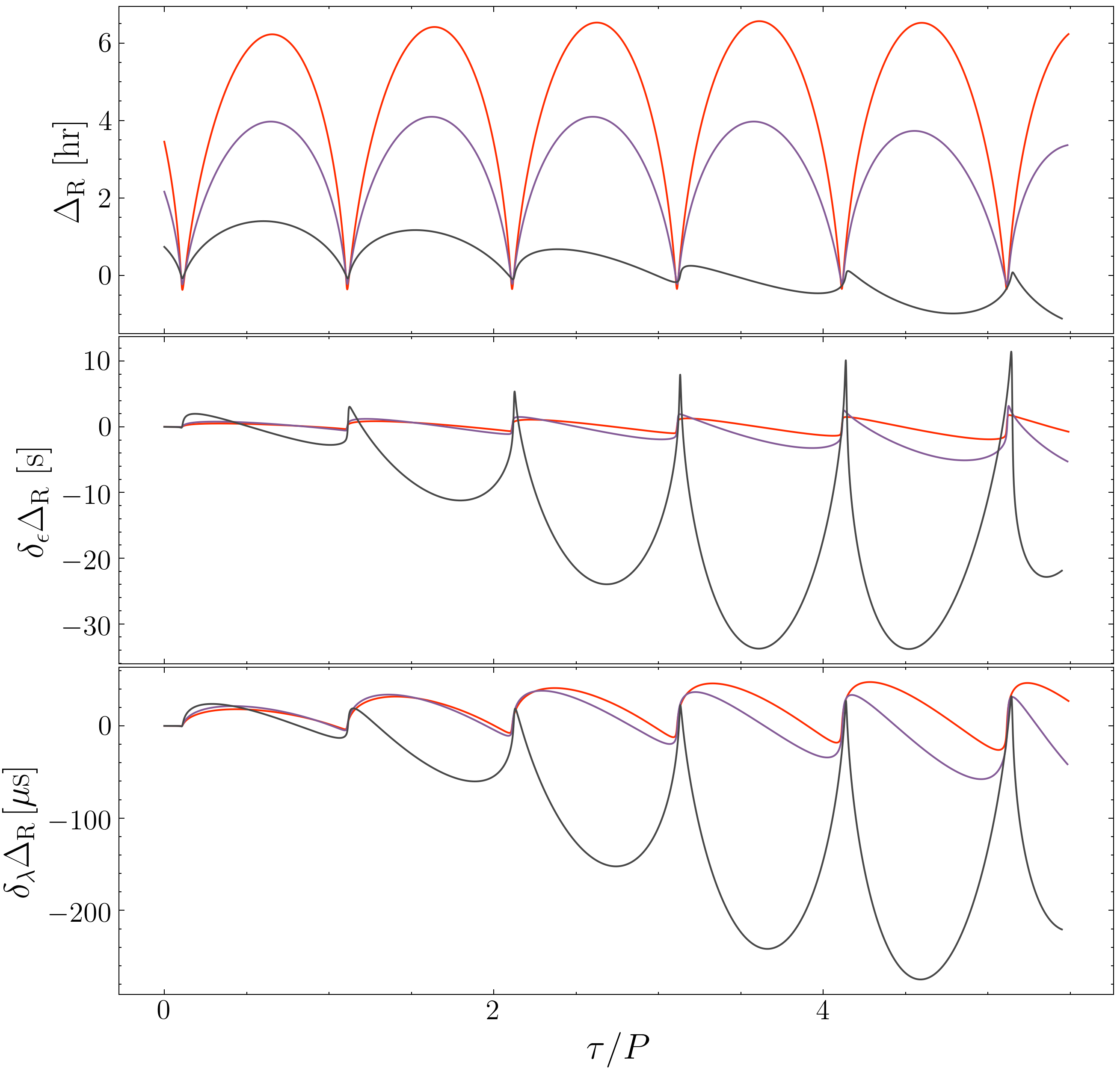}
	\caption{\textit{Top panel:} Romer Delay of a MSP orbiting the Galactic centre Sgr A* BH with $e=0.9$ and orbital period $P=0.1, 0.05,0.01$ years (red, purple, black lines respectively). The distant observer is located at $\Theta = \pi/4$. The Roemer delay varies periodically due to the eccentric orbital motion of the MSP. \textit{Middle panel:} The quadrupole-induced difference in the Roemer Delay (i.e. the timing residuals). MSPs with shorter orbital periods have greater timing residuals due to the quadrupole, but even for the longest period systems considered here the difference is of order $1$s, easily withing pulsar radio timing precisions. \textit{Bottom panel: } The timing residuals in the Roemer delay induced by spin couplings. Note the similar profile with the quadrupole-induced residuals.}
	\label{fig:Roemer}
\end{figure}

Rather than secular effects, the periodic effects induced by the quadrupole moment of the MSP have been identified as a more fruitful avenue for measuring the BH quadrupole moment. In particular, the periodic variations in the MSP coordinate position will manifest in changes in the Roemer delay of the pulsar. The Roemer delay is given by
\begin{eqnarray}
\Delta_{\rm R} = \frac{1}{c} \boldsymbol{\hat{K}} \cdot \boldsymbol{x}
\end{eqnarray}
where $\boldsymbol{\hat{K}}$ is the position unit vector of the observer and $\boldsymbol{x}$ the position vector of the MSP. Variations in the Roemer delay induced by the quadrupole will manifest in the MSP TOA residuals. The Roemer delay for a MSP-BH system is shown in Fig \ref{fig:Roemer}, along with the residuals induced by both the quadrupole moment and the MSP spin couplings. We consider a Galactic Centre-like MSP-BH system with BH mass $4.31 \times 10^6 M_{\odot}, a=+0.6$, and set the MSP to have eccentricity $e=0.9$ and consider 3 orbital periods $P = 0.1, 0.05, 0.01$ years. We set the observer at an orientation $\Theta = \pi/4$. For all these systems the Romer delay is a periodic function which varies on the scale of $\sim$ hours. The presence of a quadrupole moment $\epsilon = 0.1$ leaves the background gravitational field anisotropic and introduces a periodic timing residual. For $P = 0.1$ years, this quadrupole-induced residual $\delta_{\epsilon} \Delta_{\rm R}$ is of the order $1 $ seconds, whilst when $P=0.01$ years, $\delta_{\epsilon} \Delta_{\rm R} \sim $ is of the order of tens of seconds. Both of these residuals are well within the purview of MSP radio timing precision; the SKA is expected to enjoy timing precision in the range 10-100 ns \citep{Liu2011,Stappers2018}. In addition to the quadrupole-induced residuals, the spin couplings also introduce additional periodic variations. The potential degeneracy between these two effects will be briefly discussed later in Section \ref{sec"additionalcommnets}. These spin residuals are of the order 100's $\mu$s for $P=0.01$ years and $\sim 10 \mu$s for $P=0.1$ years. Again this is well within typical radio pulsar timing precision and so for a consistent, accurate, phase connected timing solution for precision parameter estimation of the system parameters it will be important to account for these spin effects. We have solely considered here a quadrupole moment of $\epsilon = 0.1$. If this quadrupole moment is smaller then naturally the magnitude of $\delta_{\epsilon} \Delta_{\rm R}$ will decrease whilst the magnitude of $\delta_{\lambda} \Delta_{\rm R}$ will remain unchanged. As a consequence the spin couplings could become a substantial fraction of the Roemer residuals. This again highlights the importance of a general covariant timing solution that can be applied to strong field environments for eccentric MSPs, especially if we want to use these systems for precision tests of strong-field GR; since the $\delta_{\lambda} \Delta_{\rm R}$, $\delta_{\epsilon} \Delta_{\rm R}$ follow the same general time evolution, with periodic signatures as the MSP passes through periapsis, unmodelled spin effects could imitate a non-Kerr quadrupole leading to a confusion problem for certain orbital parameters. Further, the influence of the quadrupole and spin couplings are most pronounced for close periapsis passages. Since eccentric orbits are the most desirable from the perspective of testing strong field GR, probing the quadrupole and reducing the influence of external perturbations, this is an influence that needs to be accounted for. \newline 

\begin{figure}
	\includegraphics[width=\columnwidth]{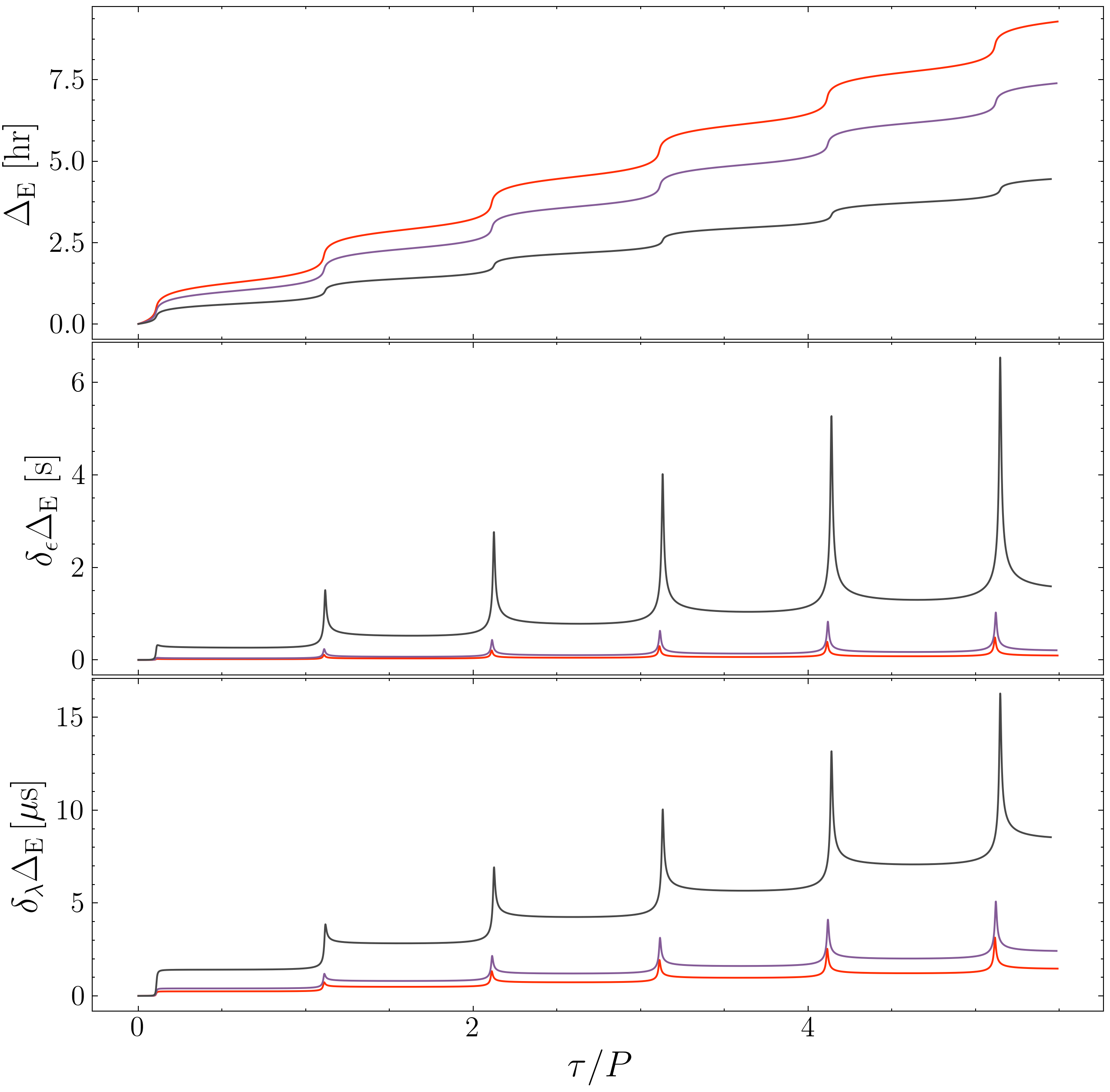}
	\caption{\textit{Top panel:} Einstein delay of a Galactic Centre MSP with orbital period P=$0.1,0.05,0.01$ years and eccentricity $e=0.9$. The Einstein delay accumulates to $\sim 9$ hrs for the longer period system and $\sim 4.5$ hrs for the shorter period system. \textit{Middle panel:} Residuals in the timing solution to the Einstein delay due to the BH quadrupole moment with $\epsilon = 0.1$. \textit{Bottom panel:}  Timing residuals in the Einstein delay due to the MSP spin couplings.}
	\label{fig:Einstein}
\end{figure}

\noindent As noted in Fig. \ref{fig:c}, due to the motion of the pulsar and the associated relativistic time dilation, pulsar signals also suffer a timing delay known as the Einstein delay, $\Delta_{\rm E}$. This delay quantifies the difference between the coordinate $t$ and proper $\tau$ times of the pulsar, i.e.
\begin{eqnarray}
\Delta_{\rm E} = t - \tau
\end{eqnarray}
Given the nature of a pulsar as a highly accurate clock, if the intrinsic rotation period of the pulsar can be established, the Einstein delay can in turn be calculated. The Einstein delay is a relativistic effect that is naturally induced through a geodesic description of the pulsar's motion. In addition, there are further contributions that arise from the quadrupole and spin. The Einstein delay for the Galactic centre systems described above is presented in Fig. \ref{fig:Einstein}. The Einstein delay accumulates over $\sim 5$ orbital phases to $\sim 4.5$ hrs for the system with $P =0.01$ years and up to $9$ hrs for the system with $P = 0.1$ years, with rapid increases as the pulsar passes through periapsis, and a general secular evolution otherwise. The quadrupole moment induces a error in the timing solution of the Einstein delay on the order of $0.1 - 1$ s, with MSP systems with shorter orbital periods most drastically affected ($\delta_{\epsilon} \Delta_{\rm E}$ peaks at $\sim 6$s for $P=0.01$ year). The error introduced due to the spin couplings is again subdominant to the quadrupole moment, of the order $1-10 \mu$s. Both the quadrupole and spin residuals follow the same general profile, with periodic rapid variations as the MSP passes through periapsis. As noted for the Roemer delay this could introduce additional complications to consistently model the MSP timing signal account for both spin couplings and the (unknown) quadrupole moment.

\subsubsection{Implications of Spin Dynamics}
Pulsar emission is not isotropic, but beamed. The misalignment between the magnetic axis $\boldsymbol{B}$ and the spin axis $\boldsymbol{S}$ is what causes the pulsed emission. The evolution of the spin axis can strongly influence the radio timing observations. Precession and nutation of the spin axis will directly influence the photon arrival times in addition to affecting the pulse profile, intensity and observed pulse frequencies \citep{Li2018,Kimpson2019B,Kocherlakota2019}. The characteristic change in the pulse frequency due to spin precession has also been suggested as a mechanism for measuring the spin parameter of the central black hole and establishing the validity or otherwise of the Cosmic  Censorship Conjecture \citep{Kocherlakota2019}. \newline 

\begin{figure}
	\includegraphics[width=\columnwidth]{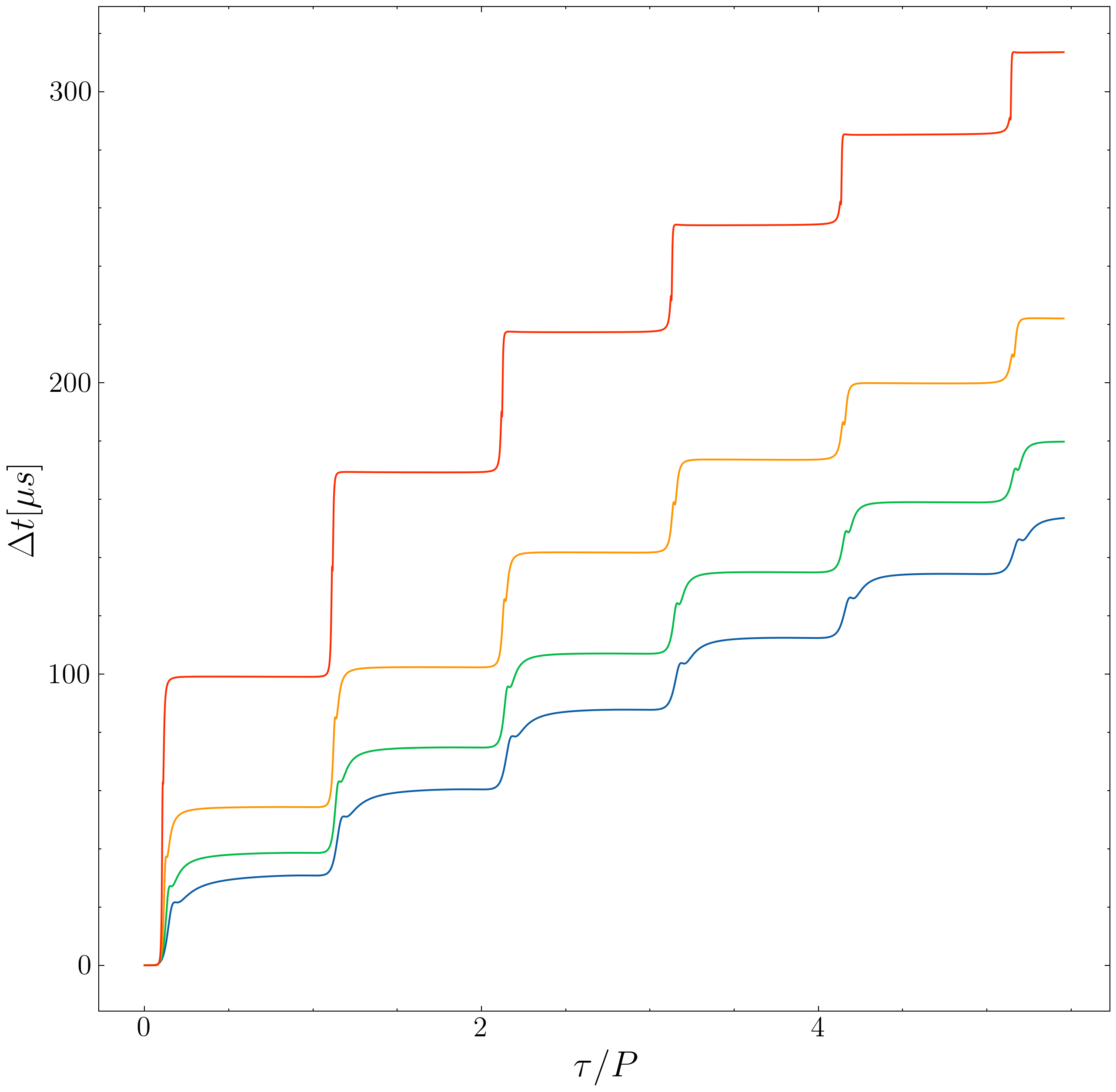}
	\caption{Timing delay due to the shift in the arrival time of the centroid of the pulse profile as a result of the precession and nutation of the MSP spin axis. The MSP has spin period $P_s = 1$ ms, semi-major axis $=200 r_{\rm g}$ and eccentricities $e=0.6,0.7,0.8,0.9$ (blue, green, orange, red lines respectively). The BH has a n intermediate mass of $M = 2.2 \times 10^3 M_{\odot}$ and spin parameter $a=0.6$. The observer is located at $\Theta = \pi/4, \Phi = 0$ and the pulsar beam is at a polar angle of $\pi/4$ with respect to the spin axis. Initially, $\theta_{\rm spin} = \pi/6$ and $\phi_{\rm spin} = 0$.}
	\label{fig:spin_time_delay}
\end{figure}

\begin{figure}
	\includegraphics[width=\columnwidth]{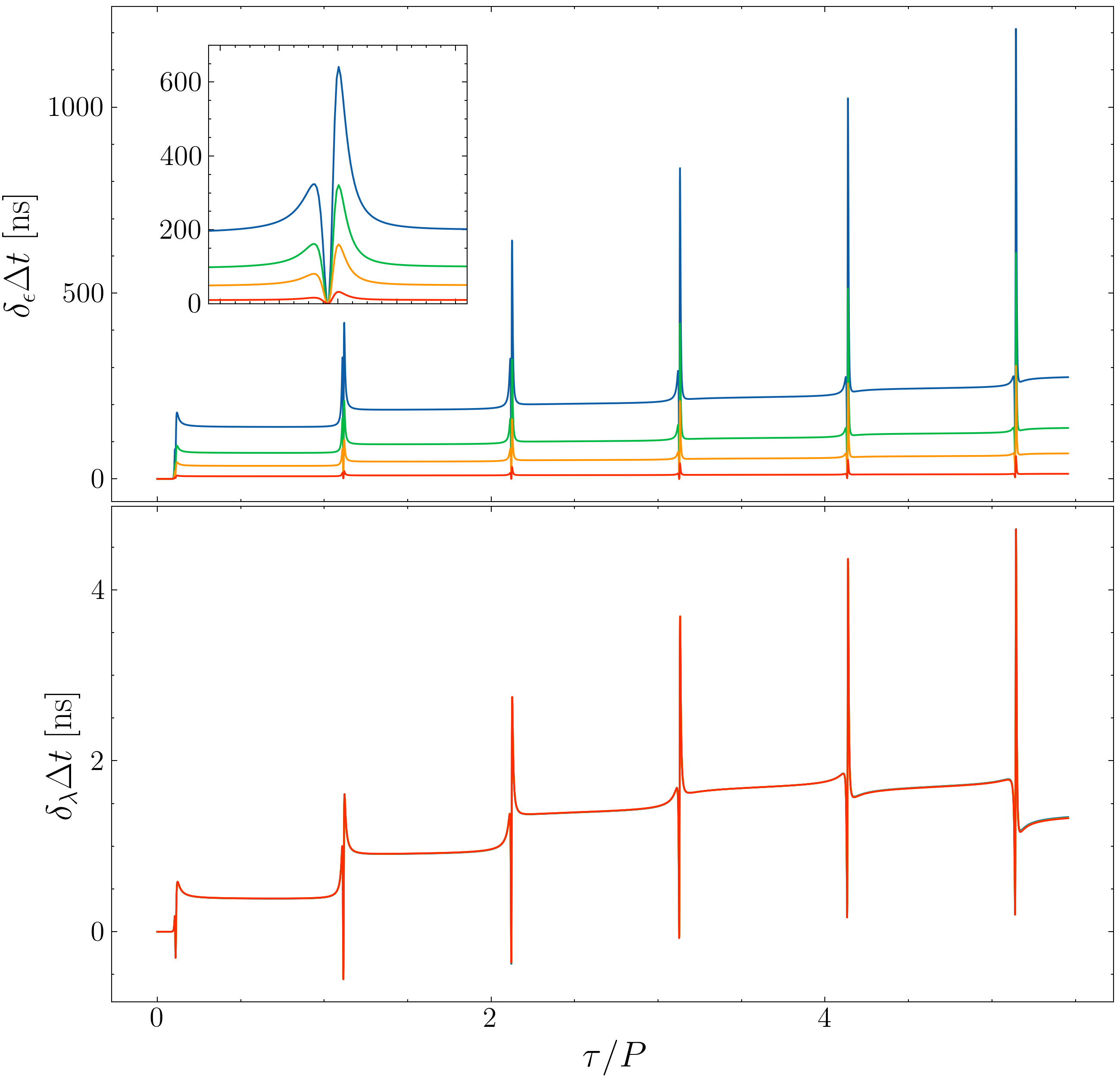}
	\caption{\textit{Top panel:} The timing residuals induced by the BH quadrupole, $\epsilon = 0.01, 0.05,0.1,0.2$ (red, orange, green ,blue respectively) for the MSP-IMBH system described above, with $e=0.9$. The additional precession and nutation induced by the BH quadrupole leads to timing delays of the order $100$'s ns, with a characteristic profile as the MSP passes through periapsis. \textit{Bottom panel:} Residuals due ot the spin couplings, which are of the order a few ns, which are likely beyond the timing precision of radio facilities. Systems with shorter orbital periods will exhibit stronger spin couplings and the associated timing delays. }
	\label{fig:spin_time_delay2}
\end{figure}

\noindent To explore the impact of the spin dynamics on the MSP radio timing we consider the evolution of the pulsar radiation axis in terms of a rotating vector model. If  $\psi, \chi(\tau)$ define the polar and azimuthal angles of the radiation beam about the pulsar spin axis (we do not consider the time evolution of the polar angle), then the evolution of the mangetic axis 3-vector is related to the spin axis as,
\begin{eqnarray}
\boldsymbol{B}(\tau)= \boldsymbol{R}_{\rm z}(\phi_{\rm spin} (\tau)) \boldsymbol{R}_{\rm y}(\theta_{\rm spin}(\tau)) 
\begin{pmatrix}
\sin \psi \cos \chi(\tau)  \\
\sin \psi \sin \chi(\tau) \\
\cos \psi
\end{pmatrix}
\end{eqnarray}
where $\boldsymbol{R}_{z,y}$ are the 3-space rotation matrices about the coordinate $z$ and $y$ axes respectively. We label the observer direction by the vector $\boldsymbol{O}$. This vector can be considered as the vector which is tangent to the asymptote that converges at the observer in a flat spacetime. The `pitch angle' $\omega$ between the radiation vector and the observer vector is then defined via,
\begin{eqnarray}
\cos \omega = \hat{\boldsymbol{B}} \cdot  \hat{\boldsymbol{O}}  
\end{eqnarray}
for unit vectors $\hat{\boldsymbol{B}}, \hat{\boldsymbol{O}}$. We define the pulse arrival time (i.e. the time centre of the pulse profile) to occur when the pitch angle is at a minimum, subject to bounds on the value of the pitch angle (the beam will not be `seen' if the pitch angle is $\pi/2$ for example). For example, if in an orthonormal basis the observer is in direction $\hat{\boldsymbol{O}} = (1,0,0)$, then the pitch angle is minimized and the centre of the pulse intersects with the observer's line of sight when the radiation vector points in the same direction, $\hat{\boldsymbol{B}} =(1,0,0) $. This gives us the condition that the pulse arrival time occurs when the beam phase $\chi(\tau)$ obtains some critical value $\chi_c$, at which $\partial_{\tau} \omega= 0$. Now, since the pulsar spin timescale ($\sim 1$ ms) is much shorter than the precession and nutation timescales of the spin axis, we can employ a two timescale approximation and neglect the evolution of $\theta_{\rm spin}, \phi_{\rm spin}$ over the MSP rotation period. If we specify that the polar angle of the radiation beam with respect to the spin axis is $\psi = \pi/4$, and the observer is at $\Theta = \pi/4, \Phi=0$ then then the critical phase angle is:

\begin{widetext}
	\begin{equation}
\chi_c = \arccos \left[ \frac{\cos \phi_{\rm spin} \cos \theta_{\rm spin} - \sin \theta_{\rm spin}}{\sqrt{\cos^2 \phi_{\rm spin} \cos^2 \theta_{\rm spin}+\sin^2 \phi_{\rm spin} + \sin^2 \theta_{\rm spin} - \cos \phi_{\rm spin} \sin 2\theta_{\rm spin} }}\right]
	\end{equation}
\end{widetext}
From this equation we can see that both the precession ($\phi_{\rm spin}$) and the nutation ($\theta_{\rm spin}$) contribute to the critical phase angle. The extra phase angle than much be traversed in order to reach the centre of the pulse profile ($\delta \chi_c$, i.e. the variation in the value and evolution of $\chi_c$ such that $\delta \chi_c =\chi_c(\tau) - \chi_c(0) $) will directly influence the observed pulse frequency. In this way, variations in the spin axis can directly imprint on the pulsar timing solution. Naturally, for variations of sufficient magnitude the spin axis variation would be so severe that the minimum of $\omega$ would be greater than the beam width and so no emission would be observed. A difference in the critical phase value is related to a timing delay as,
\begin{eqnarray}
\Delta t =  \frac{P_s}{2 \pi} \delta \chi_c
\end{eqnarray}
for MSP spin period $P_s$.  The pulse timing delay due to the time evolution of the pulsar spin axis is shown in Figure \ref{fig:spin_time_delay} for a MSP with $a_* = 200 r_{\rm g}$ around a IMBH with mass $2.2 \times 10^3 M_\odot$ \citep[the purported mass of the theoretical BH at the centre of 47 Tuc, ][]{Baumgardt2017Natur}. The timing delay due to the shift in the centre of the pulse profile is of the order 100 $\mu$s, with rapid variations as the pulsar passes through periapsis, and with greater magnitude shifts for more eccentric orbits. After $\sim 5$ orbital phases the time delay for a MSP with $e=0.9$ accumulates to $\sim 300 \mu$s. Whilst the variation in $\Delta t$ is primarily determined by the evolution of $\phi_{\rm spin}$, the change in the magnitude of the $\Delta t$ `jumps' with each periapsis passage is due to the evolution of $\theta_{\rm spin}$. 
The BH quadrupole and the MSP spin couplings can then further imprint on the pulsar timing solution, since both of these effects influence the precession and nutation of the MSP spin axis. The additional variation in the timing delay due to these effects is shown in Fig. \ref{fig:spin_time_delay2}. The presence of a non-Kerr quadrupole induces an additional timing delay as the MSP passes through periapsis, of the order of $\sim 100's$ ns. Whilst this timing delay is less than those induced by the quadrupole for e.g. the Roemer or Einstein delays, it is at the limit of MSP timing precision and the residuals also have a distinctive characteristic profile that may leave them important for real astrophysical systems. The residuals induced by the spin for the system considered here are of the order a few ns, which are unlikely to be detectable via radio timing. Naturally as the orbital radius decreases and these systems spend more time in the strong field regime the manifestation of the spin axis evolution and the contributions from the spin couplings will become more important, but we restrict ourselves here to more astrophysically likely orbital configurations.  

\begin{figure}
	\includegraphics[width=\columnwidth]{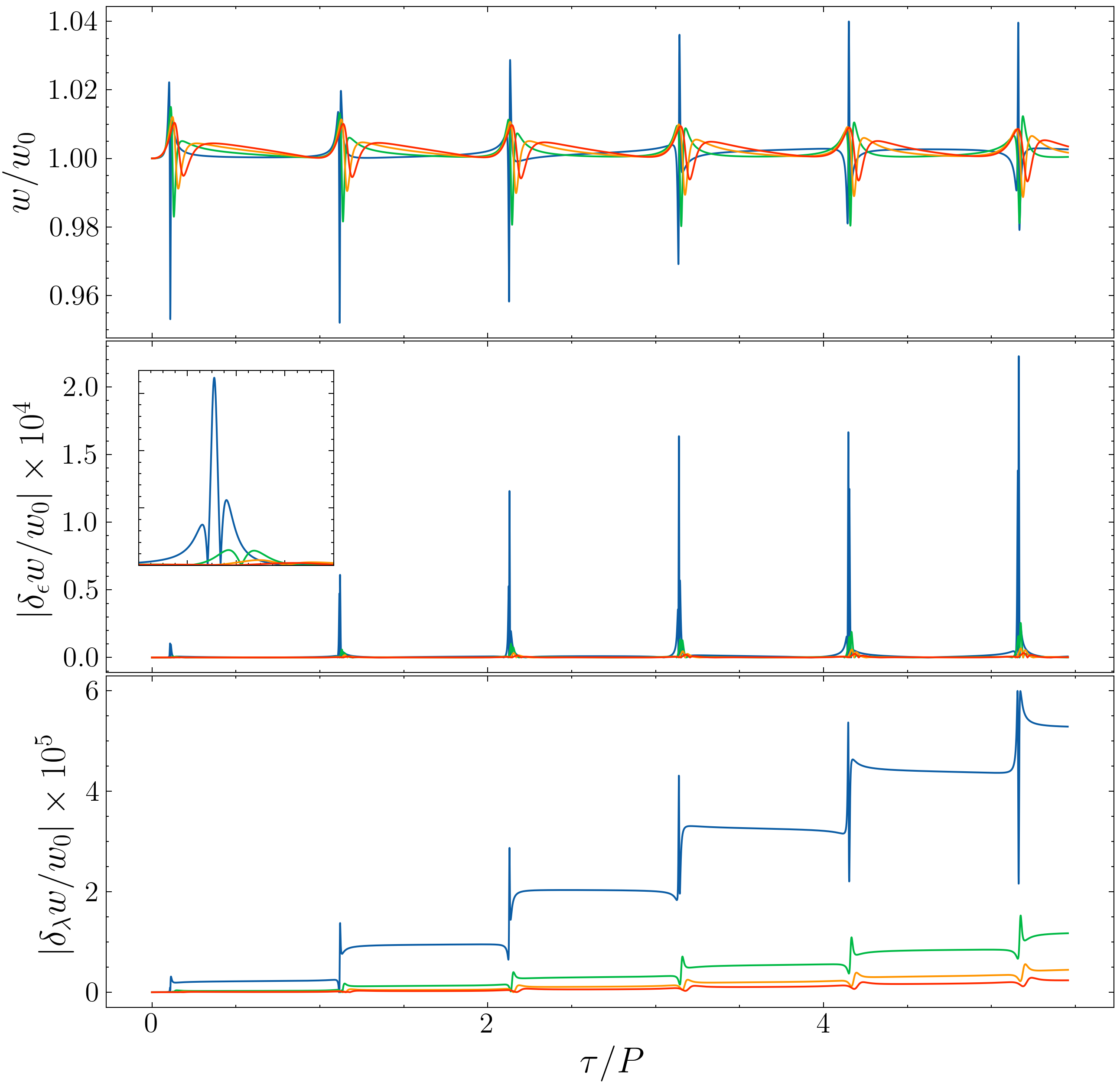}
	\caption{\textit{Top panel:} Change in the pulsar beam width for a MSP initially with $\theta_{\rm spin} = \pi/6, \phi_{\rm spin} = 0$, $\psi = \pi/12$, semi-major axis $=200 r_{\rm g}$ when in a retrograde orbit about an IMBH with $M=2.2 \times 10^3 M_{\odot}$, $a=-0.6$, a eccentricities $e = 0.9,0.8,0.7,0.6$ (blue, green, orange, red respectively). For $e=0.9$, the pulse width changes by $\sim 4 \%$, or $e=0.6$ the change is $\sim 1 \%$. \textit{Middle panel:} Variation in $w/w_0$ induced by the BH quadrupole $\epsilon = 0.1$. Rapid variations with distinctive structure are seen as the MSP passes through periapsis. \textit{Bottom panel:} Variations in $w/w_0$ induced by the MSP spin. }
	\label{fig:beam-width}
\end{figure}

Whilst the dominant contribution to $\Delta t$ is due to the spin precession, the nutation of the spin axis will also cause a shift in the observed pulse width. If the pulsar beam has half opening angle $\gamma$, then the edges of the emission cone as seen by the observer occur when $\omega = \pm \gamma$, at beam phase $\chi_1, \chi_2$, with the two roots corresponding to where the observers vector enters and leaves the pulse cross section. The angular beam width is simply,
\begin{eqnarray}
w =\chi_1 - \chi_2
\end{eqnarray}
 Since the precession does not meaningfully affect the beam width, we can set $\phi_{\rm spin}, \dot{\phi}_{\rm spin} $ and solve explicitly for the beam with:
\begin{eqnarray}
w = 2 \cos ^{-1}\left(\frac{1.41\cos\gamma  \csc\psi- \cos\theta \cot\psi - \sin\theta \cot \psi }{\cos\theta-\sin \theta}\right)
\end{eqnarray}
where we have specified the observer to be at $\Theta = \pi/4$ (a general solution is computationally straightforward to calculate, but algebraically complicated and so we do not reproduce it in full here). The evolution of the pulse width, along with the corrections induced by the quadrupole moment and the MSP spin couplings are shown in Fig. \ref{fig:beam-width}, for a MSP orbiting a 47-Tuc like IMBH, with $a=-0.6$ and semi major axis $=200 r_{\rm g}$. It can be seen that the pulse width varies due to the nutation of the spin axis on the order of $\sim 4 \%$. The corrections to the pulse width due to the $\epsilon - \lambda$ effects are smaller, of order $10^{-4}$ and $10^{-5}$ respectively for $\epsilon = 0.1$. Whilst these are small absolute numbers, the nature of MSP timing requires stacking and folding multiple pulse profiles. It is this method which leaves pulse timing so particularly sensitive; for example the pulsar PSR J0437-4715 has a spin period measured as $P_s= 5.757451924362137$ ms \citep{Verbiest2008}, which is a measurement to a precision $\sim 10^{-15}$. Consequently, even small variations in the pulse profile can prove important. 

\subsubsection{Additional comments}
\label{sec"additionalcommnets}
The timing residuals caused by the BH quadrupole and the MSP spin also raise the potential for a confusion problem: can an observer distinguish the behaviour of e.g. a non-Kerr metric with some set of orbital/pulsar parameters with the Kerr metric for some different orbital/pulsar parameters? Moreover, the $\epsilon -\lambda$ residuals exhibit similar profiles (e.g. Figs. \ref{fig:Roemer}, \ref{fig:Einstein}, \ref{fig:spin_time_delay2}), with rapid, periodic variations at periapsis. This introduces a further uncertainty for the observer: is this variation due to the spin couplings or the BH quadrupole moment? For the example systems considered in this work with $\epsilon = 0.1$, the quadrupole variations are typically large than the spin variations. In this case the spin couplings introduce an effective uncertainty into the quadrupole residuals. Self-force effects due to the mass of the MSP itself perturbing the background spacetime \citep[see][]{Barack2019} could also influence the timing signal, leading to an additional confusion source. From this it is clear that it is essential to understand and model the effect of spin couplings on pulsar ToAs in a relativistic setting. Furthermore, we have not fully explored the astrophysically relevant parameter space- instead considering just typical example systems - and for different orbital parameters or smaller values of $\epsilon$ the spin effects will become comparable. We have also not explored the influence of the BH spin parameter on the PSR timing signal \citep[e.g.][]{Zhang2017} or considered systems with a MSP and a stellar mass BH \citep[e.g. ][]{Oscoz1997}. Additionally, exploring the influence of the $ \epsilon -\lambda$ on the gravitational burst waveforms that are expected from these systems \citep[e.g.][]{Berry2013,Kimpson2020} would be an interesting further development of this work. \newline 

\noindent Moving from the time to the frequency domain, the $\epsilon - \lambda$ effects are also generally important since they intrinsic additional frequencies that must be accounted for in the Fourier analysis, particularly with regards to the evolution of the spin axis. Whilst a single isolated pulsar would exhibit one characteristic frequency set by the spin period, a MSP in the gravitational strong field would have multiple peaks in the frequency spectra. As noted in \citet{Kocherlakota2019}, this multi-peaked frequency spectra may provide a further method to extract the BH parameters.

\subsection{E/IMRI Waveform Modelling}
Waveform modelling from $\sim $stellar mass compact objects inspiraling in to much more massive BHs (Extreme/intermediate Mass Ratio Inspirals, E/IMRIs) is currently an essential research area \citep{Meent2017,Barack2019}. The detection of gravitational radiation from these systems with LISA will allow precision tests of the dynamical strong field. The eccentric systems considered in this work are particularly relevant as LISA E/IMRI sources; unlike the BH-BH binaries detected by LIGO, E/IMRIs are expected to retain significant eccentricities upon entering the LISA frequency band \citep{Amaro2018} and so accounting for this eccentricity is key to both inform the detection of these systems, and subsequent precision parameter estimation. In addition to eccentricity, the $\epsilon - \lambda$ effects will also introduce additional variations that may need to be accounted for for accurate E/IMRI waveform modelling. E/IMRI systems are expected to be observed for a large $\sim 10^4$ number of cycles at compact radii and so small perturbations can lead to significant shifts in the waveform, especially as the orbiter goes through periapsis. Whilst the dominant contributions to the waveform come from the 0th and 1st order moments, higher order effects will prove important for accurate waveform modeling; it is know for example in circular systems that spin effects leading a to a dephasing of the waveform \citep{Warburton2017}. The convolution between the eccentricity, spin and quadrupole effects and the subsequent impact on the gravitational waveform would be an interesting further study, though beyond the scope of this paper.

\section{Conclusion}
In this work we have explored the orbital-spin dynamics of a MSP in an eccentric orbit around a massive BH with an arbitrary mass quadrupole via the MPD framework. The inclusion of the BH quadrupole and the MSP spin couplings lead to perturbations in the orbital and spin evolution of the MSP. For astrphysical systems such as pulsars at the Galactic centre or the centre of globular clusters, these effects will imprint on both the pulsar timing solution (detectable with radio telescopes such as FAST, SKA, DSN) and the gravitational waveform (detectable with mHz GW detectors such as LISA). Further development of this work would be interesting, for example constructing a fully consistent, phase-connected solution to consistently model the pulsar TOAs via a relativistic pulsar timing model \citep[e.g.][]{Kimpson2019B} accounting for not just the pulsar dynamics but also the photon ray geodesic (Shapiro delay, gravitational lensing, spatial and temporal dispersion etc.) This would then allow for a much greater understanding of the observational consequences of the effects described in this work.

\bibliographystyle{mnras}
\bibliography{paper2}

\begin{thebibliography}{}
\makeatletter
\relax
\def\mn@urlcharsother{\let\do\@makeother \do\$\do\&\do\#\do\^\do\_\do\%\do\~}
\def\mn@doi{\begingroup\mn@urlcharsother \@ifnextchar [ {\mn@doi@}
  {\mn@doi@[]}}
\def\mn@doi@[#1]#2{\def\@tempa{#1}\ifx\@tempa\@empty \href
  {http://dx.doi.org/#2} {doi:#2}\else \href {http://dx.doi.org/#2} {#1}\fi
  \endgroup}
\def\mn@eprint#1#2{\mn@eprint@#1:#2::\@nil}
\def\mn@eprint@arXiv#1{\href {http://arxiv.org/abs/#1} {{\tt arXiv:#1}}}
\def\mn@eprint@dblp#1{\href {http://dblp.uni-trier.de/rec/bibtex/#1.xml}
  {dblp:#1}}
\def\mn@eprint@#1:#2:#3:#4\@nil{\def\@tempa {#1}\def\@tempb {#2}\def\@tempc
  {#3}\ifx \@tempc \@empty \let \@tempc \@tempb \let \@tempb \@tempa \fi \ifx
  \@tempb \@empty \def\@tempb {arXiv}\fi \@ifundefined
  {mn@eprint@\@tempb}{\@tempb:\@tempc}{\expandafter \expandafter \csname
  mn@eprint@\@tempb\endcsname \expandafter{\@tempc}}}

\bibitem[\protect\citeauthoryear{Abbott et~al.,}{Abbott
  et~al.}{2017}]{AbbotBHb}
Abbott B.~P.,  et~al., 2017, \mn@doi [Phys. Rev. Lett.]
  {10.1103/PhysRevLett.118.221101}, 118, 221101

\bibitem[\protect\citeauthoryear{{Amaro-Seoane}}{{Amaro-Seoane}}{2018}]{Amaro2018}
{Amaro-Seoane} P.,  2018, \mn@doi [Living Reviews in Relativity]
  {10.1007/s41114-018-0013-8}, \href
  {https://ui.adsabs.harvard.edu/abs/2018LRR....21....4A} {21, 4}

\bibitem[\protect\citeauthoryear{{Barack} \& {Pound}}{{Barack} \&
  {Pound}}{2019}]{Barack2019}
{Barack} L.,  {Pound} A.,  2019, \mn@doi [Reports on Progress in Physics]
  {10.1088/1361-6633/aae552}, \href
  {https://ui.adsabs.harvard.edu/abs/2019RPPh...82a6904B} {82, 016904}

\bibitem[\protect\citeauthoryear{{Berry} \& {Gair}}{{Berry} \&
  {Gair}}{2013}]{Berry2013}
{Berry} C.~P.~L.,  {Gair} J.~R.,  2013, \mn@doi [\mnras]
  {10.1093/mnras/sts360}, \href
  {https://ui.adsabs.harvard.edu/abs/2013MNRAS.429..589B} {429, 589}

\bibitem[\protect\citeauthoryear{{Berti}, {White}, {Maniopoulou}  \&
  {Bruni}}{{Berti} et~al.}{2005}]{Berti2005}
{Berti} E.,  {White} F.,  {Maniopoulou} A.,   {Bruni} M.,  2005, \mn@doi
  [\mnras] {10.1111/j.1365-2966.2005.08812.x}, \href
  {https://ui.adsabs.harvard.edu/abs/2005MNRAS.358..923B} {358, 923}

\bibitem[\protect\citeauthoryear{Carter}{Carter}{1968}]{Carter1968}
Carter B.,  1968, \mn@doi [Phys. Rev.] {10.1103/PhysRev.174.1559}, 174, 1559

\bibitem[\protect\citeauthoryear{{Chicone}, {Mashhoon}  \& {Punsly}}{{Chicone}
  et~al.}{2005}]{Chicone2005}
{Chicone} C.,  {Mashhoon} B.,   {Punsly} B.,  2005, \mn@doi [Physics Letters A]
  {10.1016/j.physleta.2005.05.072}, \href
  {http://adsabs.harvard.edu/abs/2005PhLA..343....1C} {343, 1}

\bibitem[\protect\citeauthoryear{{Combes}}{{Combes}}{2015}]{Combes2015}
{Combes} F.,  2015, \mn@doi [Journal of Instrumentation]
  {10.1088/1748-0221/10/09/C09001}, \href
  {https://ui.adsabs.harvard.edu/abs/2015JInst..10C9001C} {10, C09001}

\bibitem[\protect\citeauthoryear{Dixon}{Dixon}{1964}]{Dixon1964}
Dixon W.~G.,  1964, \mn@doi [Il Nuovo Cimento (1955-1965)]
  {10.1007/BF02734579}, 34, 317

\bibitem[\protect\citeauthoryear{{Dixon}}{{Dixon}}{1974}]{Dixon1974}
{Dixon} W.~G.,  1974, \mn@doi [Philosophical Transactions of the Royal Society
  of London A: Mathematical, Physical and Engineering Sciences]
  {10.1098/rsta.1974.0046}, 277, 59

\bibitem[\protect\citeauthoryear{{Event Horizon Telescope Collaboration}
  et~al.,}{{Event Horizon Telescope Collaboration} et~al.}{2019}]{EHT}
{Event Horizon Telescope Collaboration} et~al., 2019, \mn@doi [\apjl]
  {10.3847/2041-8213/ab0ec7}, \href
  {https://ui.adsabs.harvard.edu/abs/2019ApJ...875L...1E} {875, L1}

\bibitem[\protect\citeauthoryear{{Filipe Costa} \& {Nat{\'a}rio}}{{Filipe
  Costa} \& {Nat{\'a}rio}}{2014}]{Costa2014}
{Filipe Costa} L.,  {Nat{\'a}rio} J.,  2014, preprint, \href
  {http://adsabs.harvard.edu/abs/2014arXiv1410.6443F} {} (\mn@eprint {arXiv}
  {1410.6443})

\bibitem[\protect\citeauthoryear{{Glampedakis} \& {Babak}}{{Glampedakis} \&
  {Babak}}{2006}]{Glampedakis2006}
{Glampedakis} K.,  {Babak} S.,  2006, \mn@doi [Classical and Quantum Gravity]
  {10.1088/0264-9381/23/12/013}, \href
  {https://ui.adsabs.harvard.edu/abs/2006CQGra..23.4167G} {23, 4167}

\bibitem[\protect\citeauthoryear{Hui, Cheng  \& Taam}{Hui
  et~al.}{2010}]{Hui2010}
Hui C.~Y.,  Cheng K.~S.,   Taam R.~E.,  2010, \mn@doi [The Astrophysical
  Journal] {10.1088/0004-637x/714/2/1149}, 714, 1149

\bibitem[\protect\citeauthoryear{{Iorio}}{{Iorio}}{2012}]{Iorio2012}
{Iorio} L.,  2012, \mn@doi [General Relativity and Gravitation]
  {10.1007/s10714-011-1302-7}, \href
  {http://adsabs.harvard.edu/abs/2012GReGr..44..719I} {44, 719}

\bibitem[\protect\citeauthoryear{{Johannsen}}{{Johannsen}}{2013}]{Johannsen2013}
{Johannsen} T.,  2013, \mn@doi [\prd] {10.1103/PhysRevD.88.044002}, \href
  {https://ui.adsabs.harvard.edu/abs/2013PhRvD..88d4002J} {88, 044002}

\bibitem[\protect\citeauthoryear{{Kidder}}{{Kidder}}{1995}]{Kidder1995}
{Kidder} L.~E.,  1995, \mn@doi [\prd] {10.1103/PhysRevD.52.821}, \href
  {https://ui.adsabs.harvard.edu/abs/1995PhRvD..52..821K} {52, 821}

\bibitem[\protect\citeauthoryear{{Kimpson}, {Wu}  \& {Zane}}{{Kimpson}
  et~al.}{2019}]{Kimpson2019B}
{Kimpson} T.,  {Wu} K.,   {Zane} S.,  2019, \mn@doi [\mnras]
  {10.1093/mnras/stz845}, \href
  {https://ui.adsabs.harvard.edu/abs/2019MNRAS.486..360K} {486, 360}

\bibitem[\protect\citeauthoryear{{Kimpson}, {Wu}  \& {Zane}}{{Kimpson}
  et~al.}{2020}]{Kimpson2020}
{Kimpson} T.,  {Wu} K.,   {Zane} S.,  2020, \mn@doi [\mnras]
  {10.1093/mnras/staa1259}, \href
  {https://ui.adsabs.harvard.edu/abs/2020MNRAS.495..600K} {495, 600}

\bibitem[\protect\citeauthoryear{{K{\i}z{\i}ltan}, {Baumgardt}  \&
  {Loeb}}{{K{\i}z{\i}ltan} et~al.}{2017}]{Baumgardt2017Natur}
{K{\i}z{\i}ltan} B.,  {Baumgardt} H.,   {Loeb} A.,  2017, \mn@doi [\nat]
  {10.1038/nature21361}, \href
  {https://ui.adsabs.harvard.edu/abs/2017Natur.542..203K} {542, 203}

\bibitem[\protect\citeauthoryear{{Kocherlakota}, {Joshi}, {Bhattacharyya},
  {Chakraborty}, {Ray}  \& {Biswas}}{{Kocherlakota}
  et~al.}{2019}]{Kocherlakota2019}
{Kocherlakota} P.,  {Joshi} P.~S.,  {Bhattacharyya} S.,  {Chakraborty} C.,
  {Ray} A.,   {Biswas} S.,  2019, \mn@doi [\mnras] {10.1093/mnras/stz2538},
  \href {https://ui.adsabs.harvard.edu/abs/2019MNRAS.490.3262K} {490, 3262}

\bibitem[\protect\citeauthoryear{{Li}, {Wu}  \& {Singh}}{{Li}
  et~al.}{2019}]{Li2018}
{Li} K.~J.,  {Wu} K.,   {Singh} D.,  2019, \mn@doi [\mnras]
  {10.1093/mnras/stz389}, \href
  {https://ui.adsabs.harvard.edu/abs/2019MNRAS.485.1053L} {485, 1053}

\bibitem[\protect\citeauthoryear{{Liu}, {Verbiest}, {Kramer}, {Stappers}, {van
  Straten}  \& {Cordes}}{{Liu} et~al.}{2011}]{Liu2011}
{Liu} K.,  {Verbiest} J.~P.~W.,  {Kramer} M.,  {Stappers} B.~W.,  {van Straten}
  W.,   {Cordes} J.~M.,  2011, \mn@doi [\mnras]
  {10.1111/j.1365-2966.2011.19452.x}, \href
  {https://ui.adsabs.harvard.edu/abs/2011MNRAS.417.2916L} {417, 2916}

\bibitem[\protect\citeauthoryear{Liu, Wex, Kramer, Cordes  \& Lazio}{Liu
  et~al.}{2012}]{Liu2012}
Liu K.,  Wex N.,  Kramer M.,  Cordes J.~M.,   Lazio T. J.~W.,  2012, Astrophys.
  J., 747

\bibitem[\protect\citeauthoryear{{Lorimer}}{{Lorimer}}{2008}]{Lorimer2008}
{Lorimer} D.~R.,  2008, \mn@doi [Living Rev. Relativ.] {10.12942/lrr-2008-8},
  \href {http://ukads.nottingham.ac.uk/abs/2008LRR....11....8L} {11, 8}

\bibitem[\protect\citeauthoryear{{Majid}, {Prince}, {Pearlman}, {Kocz}  \&
  {Horiuchi}}{{Majid} et~al.}{2019}]{Majid2019}
{Majid} W.~A.,  {Prince} T.~A.,  {Pearlman} A.~B.,  {Kocz} J.,   {Horiuchi} S.,
   2019, in AAS/High Energy Astrophysics Division. AAS/High Energy Astrophysics
  Division.
p. 112.80

\bibitem[\protect\citeauthoryear{{Mashhoon} \& {Singh}}{{Mashhoon} \&
  {Singh}}{2006}]{Mashhoon2006}
{Mashhoon} B.,  {Singh} D.,  2006, \mn@doi [\prd] {10.1103/PhysRevD.74.124006},
  \href {http://adsabs.harvard.edu/abs/2006PhRvD..74l4006M} {74, 124006}

\bibitem[\protect\citeauthoryear{{Mathisson}}{{Mathisson}}{1937}]{Mathisson1937}
{Mathisson} A.,  1937, Acta Phys. Pol., 6, 163

\bibitem[\protect\citeauthoryear{{O'Connell}}{{O'Connell}}{1969}]{Connell1969}
{O'Connell} R.~F.,  1969, \mn@doi [\apss] {10.1007/BF00651266}, \href
  {https://ui.adsabs.harvard.edu/abs/1969Ap&SS...4..119O} {4, 119}

\bibitem[\protect\citeauthoryear{{Oscoz}, {Goicoechea}, {Mediavilla}  \&
  {Buitrago}}{{Oscoz} et~al.}{1997}]{Oscoz1997}
{Oscoz} A.,  {Goicoechea} L.~J.,  {Mediavilla} E.,   {Buitrago} J.,  1997,
  \mn@doi [\mnras] {10.1093/mnras/285.2.413}, \href
  {https://ui.adsabs.harvard.edu/abs/1997MNRAS.285..413O} {285, 413}

\bibitem[\protect\citeauthoryear{Pan, Hobbs, Li, Ridolfi, Wang  \& Freire}{Pan
  et~al.}{2016}]{Pan2016}
Pan Z.,  Hobbs G.,  Li D.,  Ridolfi A.,  Wang P.,   Freire P.,  2016, \mn@doi
  [Monthly Notices of the Royal Astronomical Society: Letters]
  {10.1093/mnrasl/slw037}, 459, L26

\bibitem[\protect\citeauthoryear{Papapetrou}{Papapetrou}{1951}]{Papapetrou1951}
Papapetrou A.,  1951, \mn@doi [Proceedings of the Royal Society of London A:
  Mathematical, Physical and Engineering Sciences] {10.1098/rspa.1951.0200},
  209, 248

\bibitem[\protect\citeauthoryear{{Pearlman}, {Majid}, {Prince}, {Kocz},
  {Horiuchi}  \& {Naudet}}{{Pearlman} et~al.}{2019}]{Pearlman2019}
{Pearlman} A.~B.,  {Majid} W.~A.,  {Prince} T.~A.,  {Kocz} J.,  {Horiuchi} S.,
   {Naudet} C.~J.,  2019, in AAS/High Energy Astrophysics Division. AAS/High
  Energy Astrophysics Division.
p. 112.105

\bibitem[\protect\citeauthoryear{{Plyatsko} \& {Fenyk}}{{Plyatsko} \&
  {Fenyk}}{2016}]{Plyatsko2016}
{Plyatsko} R.,  {Fenyk} M.,  2016, \mn@doi [\prd] {10.1103/PhysRevD.94.044047},
  \href {http://adsabs.harvard.edu/abs/2016PhRvD..94d4047P} {94, 044047}

\bibitem[\protect\citeauthoryear{{Psaltis}}{{Psaltis}}{2008}]{Psaltis2008}
{Psaltis} D.,  2008, \mn@doi [Living Reviews in Relativity]
  {10.12942/lrr-2008-9}, \href
  {http://adsabs.harvard.edu/abs/2008LRR....11....9P} {11, 9}

\bibitem[\protect\citeauthoryear{{Rajwade}, {Lorimer}  \& {Anderson}}{{Rajwade}
  et~al.}{2017}]{Rajwade2017}
{Rajwade} K.~M.,  {Lorimer} D.~R.,   {Anderson} L.~D.,  2017, \mn@doi [Mon.
  Not. R. Astron. Soc.] {10.1093/mnras/stx1661}, \href
  {http://ukads.nottingham.ac.uk/abs/2017MNRAS.471..730R} {471, 730}

\bibitem[\protect\citeauthoryear{{Saxton}, {Younsi}  \& {Wu}}{{Saxton}
  et~al.}{2016}]{Saxton2016}
{Saxton} C.~J.,  {Younsi} Z.,   {Wu} K.,  2016, \mn@doi [Mon. Not. R. Astron.
  Soc.] {10.1093/mnras/stw1626}, \href
  {http://ukads.nottingham.ac.uk/abs/2016MNRAS.461.4295S} {461, 4295}

\bibitem[\protect\citeauthoryear{{Schmidt}}{{Schmidt}}{2002}]{Schmidt2002}
{Schmidt} W.,  2002, \mn@doi [Classical and Quantum Gravity]
  {10.1088/0264-9381/19/10/314}, \href
  {https://ui.adsabs.harvard.edu/abs/2002CQGra..19.2743S} {19, 2743}

\bibitem[\protect\citeauthoryear{{Singh}}{{Singh}}{2005}]{Singh2005}
{Singh} D.,  2005, \mn@doi [\prd] {10.1103/PhysRevD.72.084033}, \href
  {http://adsabs.harvard.edu/abs/2005PhRvD..72h4033S} {72, 084033}

\bibitem[\protect\citeauthoryear{{Singh}, {Wu}  \& {Sarty}}{{Singh}
  et~al.}{2014}]{Singh2014}
{Singh} D.,  {Wu} K.,   {Sarty} G.~E.,  2014, \mn@doi [Mon. Not. R. Astron.
  Soc.] {10.1093/mnras/stu614}, \href
  {http://ukads.nottingham.ac.uk/abs/2014MNRAS.441..800S} {441, 800}

\bibitem[\protect\citeauthoryear{{Stappers}, {Keane}, {Kramer}, {Possenti}  \&
  {Stairs}}{{Stappers} et~al.}{2018}]{Stappers2018}
{Stappers} B.~W.,  {Keane} E.~F.,  {Kramer} M.,  {Possenti} A.,   {Stairs}
  I.~H.,  2018, \mn@doi [Philosophical Transactions of the Royal Society of
  London Series A] {10.1098/rsta.2017.0293}, \href
  {https://ui.adsabs.harvard.edu/abs/2018RSPTA.37670293S} {376, 20170293}

\bibitem[\protect\citeauthoryear{{Tulczyjew}}{{Tulczyjew}}{1959}]{Tulczyjew1959}
{Tulczyjew} W.,  1959, \mn@doi [Acta Phys. Pol.] {10.1093/mnras/stu614}, 18,
  393

\bibitem[\protect\citeauthoryear{{Verbiest} et~al.,}{{Verbiest}
  et~al.}{2008}]{Verbiest2008}
{Verbiest} J.~P.~W.,  et~al., 2008, \mn@doi [\apj] {10.1086/529576}, \href
  {https://ui.adsabs.harvard.edu/abs/2008ApJ...679..675V} {679, 675}

\bibitem[\protect\citeauthoryear{{Verbiest} et~al.,}{{Verbiest}
  et~al.}{2009}]{Verbiest2009}
{Verbiest} J.~P.~W.,  et~al., 2009, \mn@doi [\mnras]
  {10.1111/j.1365-2966.2009.15508.x}, \href
  {http://adsabs.harvard.edu/abs/2009MNRAS.400..951V} {400, 951}

\bibitem[\protect\citeauthoryear{{Warburton}, {Osburn}  \& {Evans}}{{Warburton}
  et~al.}{2017}]{Warburton2017}
{Warburton} N.,  {Osburn} T.,   {Evans} C.~R.,  2017, \mn@doi [\prd]
  {10.1103/PhysRevD.96.084057}, \href
  {https://ui.adsabs.harvard.edu/abs/2017PhRvD..96h4057W} {96, 084057}

\bibitem[\protect\citeauthoryear{{Wex} \& {Kopeikin}}{{Wex} \&
  {Kopeikin}}{1999}]{Wex1999}
{Wex} N.,  {Kopeikin} S.~M.,  1999, \mn@doi [\apj] {10.1086/306933}, \href
  {https://ui.adsabs.harvard.edu/abs/1999ApJ...514..388W} {514, 388}

\bibitem[\protect\citeauthoryear{Wharton, Chatterjee, Cordes, Deneva  \&
  Lazio}{Wharton et~al.}{2012}]{Wharton2012}
Wharton R.~S.,  Chatterjee S.,  Cordes J.~M.,  Deneva J.~S.,   Lazio T. J.~W.,
  2012, \mn@doi [Astrophys. J.] {10.1088/0004-637X/753/2/108}, 753

\bibitem[\protect\citeauthoryear{{Will}}{{Will}}{2014}]{Will2014}
{Will} C.~M.,  2014, \mn@doi [Living Rev. Relativ.] {10.12942/lrr-2014-4},
  \href {http://ukads.nottingham.ac.uk/abs/2014LRR....17....4W} {17, 4}

\bibitem[\protect\citeauthoryear{{Zhang} \& {Saha}}{{Zhang} \&
  {Saha}}{2017}]{Zhang2017}
{Zhang} F.,  {Saha} P.,  2017, \mn@doi [\apj] {10.3847/1538-4357/aa8f47}, \href
  {https://ui.adsabs.harvard.edu/abs/2017ApJ...849...33Z} {849, 33}

\bibitem[\protect\citeauthoryear{{van de Meent}}{{van de
  Meent}}{2017}]{Meent2017}
{van de Meent} M.,  2017, in Journal of Physics Conference Series. p. 012022,
  \mn@doi{10.1088/1742-6596/840/1/012022}

\makeatother
\end{thebibliography}

% Don't change these lines
\bsp	% typesetting comment
\label{lastpage}
\end{document}